\documentclass[%
 reprint,
superscriptaddress,
showpacs,preprintnumbers,
 amsmath,amssymb,
 aps,
 prb,
]{revtex4-2}
\usepackage{graphicx}
\usepackage{xcolor}
\usepackage{dcolumn}
\usepackage{bm}

\begin{document}

\title{Model for the commensurate charge-density waves in under-hole-doped cuprate superconductors}

\author{Jacob Morrissey}
\affiliation{Department of Physics, University of Central Florida, Orlando, FL 32816-2385, USA}
\author{Timothy J. Haugan}
\affiliation{U. S. Air Force Research Laboratory, Wright-Patterson Air Force Base, Ohio 45433-7251, USA}
\author{Richard A. Klemm}
\email{richard.klemm@ucf.edu, corresponding author}
\affiliation{Department of Physics, University of Central Florida, Orlando, FL 32816-2385, USA}
\affiliation{U. S. Air Force Research Laboratory, Wright-Patterson Air Force Base, Ohio 45433-7251, USA}
\date{\today}

\begin{abstract}
A simple model of the  commensurate charge-density wave (CCDW) portion of the underdoped pseudogap regions of monolayer Bi$_2$Sr$_{2-x}$La$_x$CuO$_{6-x}$ (Bi2201), bilayer   Bi$_2$Sr$_2$CaCu$_2$O$_{8+\delta}$ (Bi2212), and trilayer Bi$_2$Sr$_2$Ca$_2$Cu$_3$O$_{10+\delta}$ (Bi2223) cuprate superconductors is presented and studied. Above the superconducting transition temperature $T_c$ but below the pseudogap transition temperature $T_p > T_c$,  the CCDW forms on the oxygen sites in the CuO$_2$  layers with excess charges of $\pm\delta e$, where $e$ is the electronic charge, forming on alternating oxygen sites.  This model is equivalent to $N$-layer versions of the two-dimensional Ising model for spins on a square lattice with repulsive interactions $J' , J>0$ between near-neighbor inter- and intralayer sites, respectively.  For strong coupling, we show analytically for sections of $L\times M\times N$ sites that the partition function  in the  $J'\rightarrow\pm \infty$ limits reduces to that for an effective single layer with $L\times M$ sites and $J$ replaced by $NJ$.  The CCDW is therefore strongly enhanced and stabilized by  multilayer structures, likely accounting for the enhanced THz emission observed from the intrinsic Josephson junctions in underdoped  Bi2212 mesas and for the many experiments on Bi2212 and related compounds purporting to provide evidence for a superconducting order parameter with $d_{x^2-y^2}$-wave symmetry.
\end{abstract}

\pacs{} \vskip0pt
\maketitle

\section{Introduction}
Layered materials that are metallic at high temperatures are susceptible to phase transitions at lower temperatures into a variety of states including superconductivity, charge-density waves (CDWs), and spin-density waves \cite{book,Klemm1,TMDsuperconductors}.  One of the first examples of a class of materials to exhibit this complexity was the transition metal dichalcogenides \cite{WilsonDiSalvoMahajan}.  The first example of these materials, $2H$-TaS$_2$,  exhibits a strongly anisotropic resistivity at high temperatures that is extremely linear in the temperature $T$ parallel to the layers, but exhibits a weakly first-order phase transition at about 75 K, below which its resistivities parallel and perpendicular to the layers exhibit strongly different temperature dependencies, quasi-metallic and quasi-insulating, respectively, followed by a a bulk phase transition into the superconducting state at about 0.6 K \cite{book,Frindt}. X-ray analysis determined that the state between 0.6 K and 75 K was a commensurate charge-density wave (CCDW) \cite{WilsonDiSalvoMahajan}.  Angle-resolved photoemission (ARPES) studies of  $2H$-TaS$_2$ indicated that it could exhibit  nodes at particular points in its first Brillouin zone \cite{Tonjes}.  Intercalation with pyridine destroyed the CCDW and raised the 
superconducting transition to about 3.6 K \cite{book, GDKG}, indicating both that the CCDW is stabilized by interlayer interactions such as the Coulomb interaction, and that it competes with  the superconductivity.  Hall effect measurements on pristine $2H$-TaS$_2$ indicated that the CCDW in that material formed on one or more intralayer  hole bands, as shown in Fig. 1 \cite{book,Thompson}.  Superconducting fluctuation effects allowed for measurements up to temperatures on the order of 3 $T_c$. \cite{Klemm3,KLB}

 \begin{figure}
\center{\includegraphics[width=0.35\textwidth]{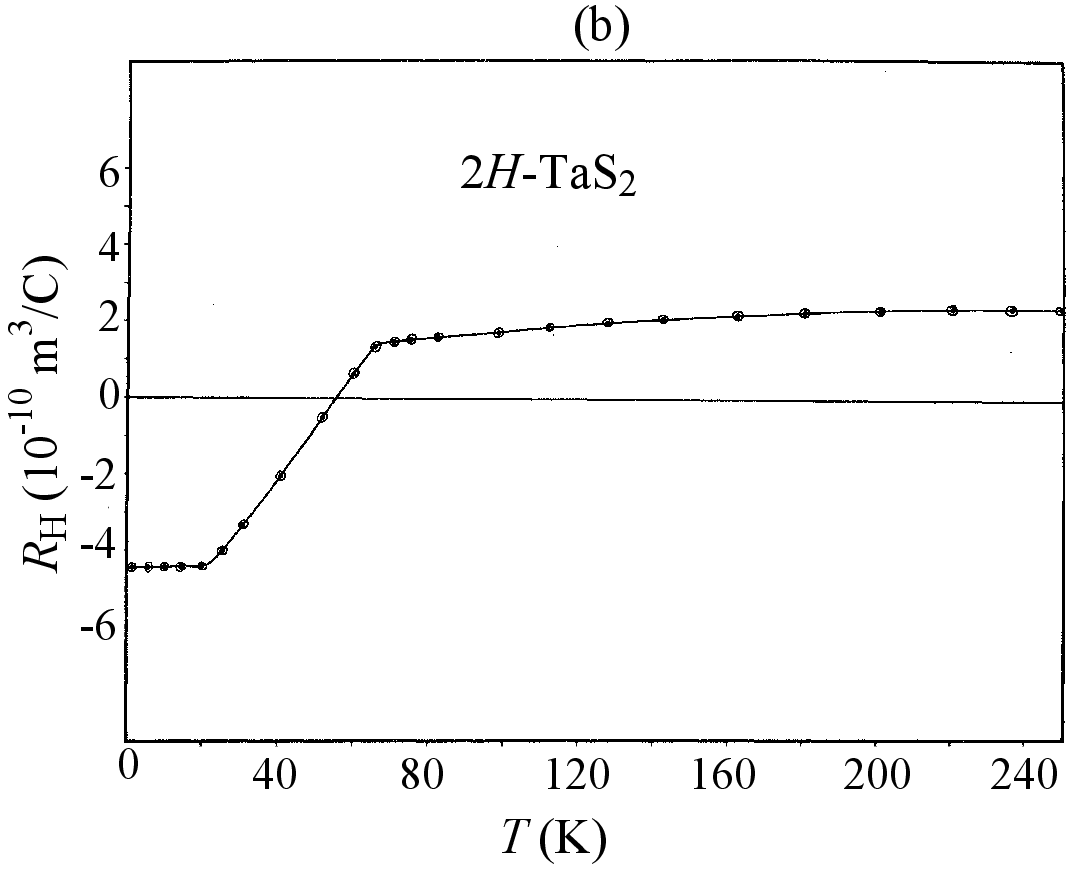}
\caption{Hall constant $R_H$ versus temperature $T$ for transport parallel to the layers of  2$H$-TaS$_2$.  \cite{book,Thompson}}}
\end{figure}

There are two primary objectives of this  work.  The first is to improve the THz emission power from the intrinsic Josephson junctions in the high-temperature layered superconductor  Bi$_2$Sr$_2$CaCu$_2$O$_{8+\delta}$ (Bi2212).  It has generally been known from many laboratories that have worked on  this topic that the most powerful emission arises from underdoped Bi2212 samples.  However, underdoped  Bi$_2$Sr$_{2-x}$La$_x$CuO$_{6-x}$ (Bi2201), Bi2212, and  Bi$_2$Sr$_2$Ca$_2$Cu$_3$O$_{10+\delta}$ (Bi2223) exhibit what is often characterized as a ``pseudogap'', the nature of which has not been well understood, except that  its onset temperature $T_p$ is generally higher than  the superconducting transition temperature $T_c$ over a substantial but incomplete region of the  hole-doping phase diagram dependent upon the oxygen non-stoichiometry concentration $\delta$.

 It has has long been known in the transition metal dichalcogenides that there is a competition between the superconducting order parameter and the CDW order parameter, which can be either incommensurate with the crystal lattice or commensurate with it \cite{book,WilsonDiSalvoMahajan}.  However, what has not been generally appreciated is that the CDW forms on one or more hole bands for transport parallel to the layers, as indicated by the temperature $T$ dependence of the Hall constant $R_H$ parallel to the layers in 2$H$-TaS$_2$, the data of which are shown in Fig. 1.  Assuming the pseudogap in Bi2212 is due to  a CDW,  we further assume that in order to strengthen the output power of the THz emission, it must be a CCDW.  Otherwise, under intense emission activity in the current-voltage phase diagram, an incommensurate CDW (ICDW), most likely generated by intralayer Fermi surface nesting,  could move about in the relevant layers,  interfering with the emission and reducing its output power.

The second objective is to try to help resolve the
long-standing raging controversy over the orbital symmetry of the superconducting order parameter in the high transition temperature $T_c$ cuprate superconductors \cite{VanHarlingen,TsueiKirtley1,TsueiKirtley2,KRS1,KARS,Li,KRS2,BKS,Takano1,Takano2,Takano3,Takano4,Klemm2,Latyshev,KlemmPhilMag,Kadowaki,Zhong,KivelsonSpivak,Zhu,Xue,JuHongLee,Zhao1,ZXShen,electron-doped,Poccia,Kim,Tsinghuapublished}.  Early ``phase-sensitive'' experiments, that purported to be sensitive to the phase of the superconducting order parameter,  on the two-layer cuprate YBa$_2$Cu$_3$O$_{7-\delta}$ (YBCO), provided evidence that was claimed to be consistent with  $d_{x^2-y^2}$-wave symmetry of the superconducting order parameter\cite{VanHarlingen,TsueiKirtley1,TsueiKirtley2}.  However, after preliminary theoretical studies \cite{KRS1,KARS}, the phase-sensitive $c$-axis twist experiment on the overdoped two-layer cuprate superconductor Bi2212 \cite{Li}, and its theoretical support \cite{KRS2, BKS}, contradicted those conclusions for YBCO.  Since there was a desire for experiments on smaller samples to be performed, a series of experiments on artificial Bi2212 cross-whiskers led to somewhat ambiguous results \cite{Takano1,Takano2,Takano3,Takano4}, which were explained as being inconclusive \cite{Klemm2}, and later contradicted by experiments on naturally-formed Bi2212 cross-whiskers \cite{Latyshev}.  Those experiments and theories were summarized in a review article \cite{KlemmPhilMag}.  In Bi2212, there is a wide range of hole-doping compositions, leading to a large variety of point contact scanning tunneling microscopy (STM) results.  Among many suggestions for complicated $d$-wave order parameter spatial configurations, one suggested a wire of $d$-wave constituents and random distributions that would average spatially to zero \cite{KivelsonSpivak}.

  Meanwhile, several workers who found STM evidence for an isotropic $s$-wave superconducting density of states in Bi2212 were having extreme difficulty in getting their results published, but at least two groups managed to do so \cite{Kadowaki,Zhong}, and their results are shown in Figs. 2 and 3.
\begin{figure}
\center{\includegraphics[width=0.3\textwidth]{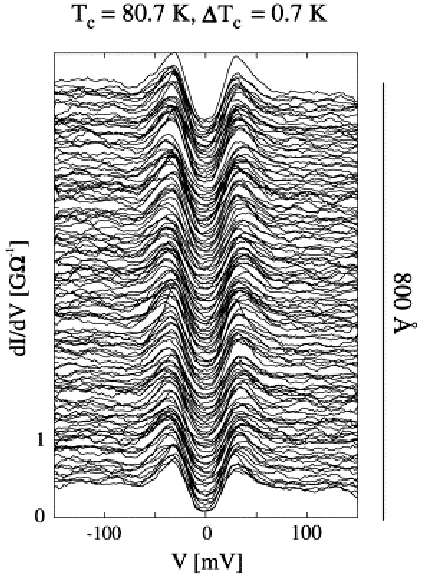}
\caption{Superconducting gap of  Bi2212 measured by STM at 4.2K in vacuum.  Reprinted with permission from  B. W. Hoogenboom, K. Kadowaki, B. Revaz, and {\O}. Fischer, Homogeneous samples of Bi$_2$Sr$_2$CaCu$_2$O$_{8+\delta}$, Physica C {\bf 391}, 376-380 (2003). \cite{Kadowaki}}}
\end{figure}
\begin{figure}
\center{\includegraphics[width=0.3\textwidth]{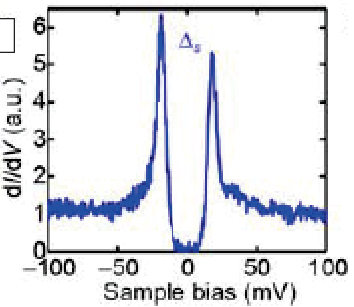}
\caption{Superconducting gap of  Bi2212 covered with a monolayer of CuO$_2$ measured at 4.2K. Reprinted with permission from  Y. Zhong, Y. Wang, S. Han, Y.-R. Lv, W.-L. Wang, D. Zhang, H. Ding, Y.-M. Zhang, L. Wang, K. He, R. Zhong, H A. Schneeloch, G. D. Gu, C.-L. Song, X.-C. Ma, and Q. K. Xue, Nodeless pairing in superconducting copper-oxide monolayer films on Bi$_2$Sr$_2$CaCu$_2$O$_{8+\delta}$, Sci. Bull. {\bf 61}, 1239 (2016). \cite{Zhong}}}
\end{figure}

Two decades after the first ``phase-sensitive'' experiments, the breakthrough came from the Tsinghua University group in Beijing.   They freshly cleaved Bi2212 in high vacuum, deposited a monolayer of CuO$_2$ on its top, and performed STM on it \cite{Zhong}.  They found that the Bi2212 sample was an inhomogeneous mixture of two distinct nanodomains, as shown in Fig. 4.  One type of these nanodomains  characterizes the ``pseudogap'' regions, and the other type characterizes the superconducting domains.   The pseudogap has a nodal density of states (with a nearly V-shaped center, consistent with a pseudogap order parameter with line nodes) that disappears at $T_p>T_c$, as pictured in the right panel of Fig. 4. In Fig. 3 and in the left panel of Fig. 4, the low-temperature superconducting density of states they found is entirely consistent with an isotropic, $s$-wave  density of states, as in the Bardeen-Cooper-Schrieffer (BCS) theory \cite{BCS}, 
\begin{eqnarray}
\rho_{BCS}(E)&=&\frac{|E|\Theta[E^2-\Delta^2(0)]}{\sqrt{E^2-\Delta^2(0)}},\label{BCSDOS}
\end{eqnarray}
 where $\Theta(x)$ is the Heaviside step function and $\Delta(0)$ is the superconducting gap at $T=0$.  It is symmetric about $E=0$, as in Figs. 2 and 3.  We note that the superconducting density of states of a randomly distributed $d$-wave superconductor as in Ref. \cite{KivelsonSpivak}, with a planar order parameter $\Delta(\theta)=\Delta_0\cos(2\theta)$, where $\theta$ is a random variable, would be identical to that of a spatially ordered $d$-wave superconductor, which is
\begin{eqnarray}
\rho_{d_{x^2-y^2}}(E)&=&\frac{2E}{\pi}\int_0^{\pi/2}\frac{d\theta\Theta[E-\Delta(\theta)]}{\sqrt{E^2-\Delta^2(\theta)}},\label{dwaveDOS}\\
&=&\frac{2}{\pi}\Big\{(E/\Delta_0)K(E/\Delta_0)\Theta(\Delta_0-E)\nonumber\\
& &+K(\Delta_0/E)\Theta(E-\Delta_0)\Big\},
\end{eqnarray}
where $K(x)$ is the complete elliptic integral of the first kind \cite{Mahan}, which is linear in $x$ for small $x$ and diverges logarithmically at $x=1$. We note that the energy gained from a normal metal transforming into the isotropic BCS $s$-wave superconducting state is much greater than the energy gained by a normal metal transforming into a $d$-wave superconducting state, implying that the latter scenario is unlikely to occur in nature, except possibly in highly unusual circumstances, such as those forbidding an $s$-wave superconducting state altogether or based upon a non-phonon exchange pairing mechanism.  For example, triplet spin pairing favors one or more types of $p$-wave states \cite{SK1,SK2}.
 
\begin{figure}
\center{\includegraphics[width=0.49\textwidth]{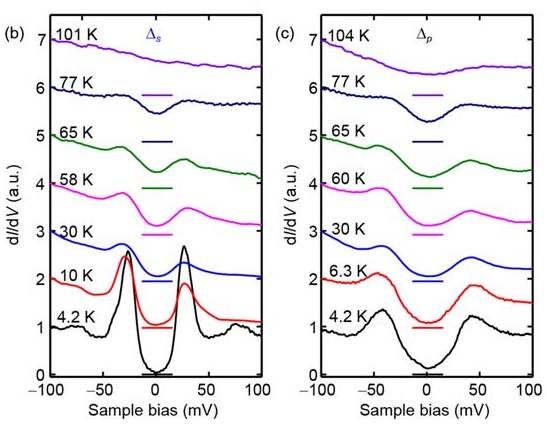}
\caption{Temperature $T$ dependencies of the superconducting gap (left) and the  pseudogap (right) of Bi2212. The low-$T$ superconducting gap is of the classic Bardeen-Cooper-Schrieffer (BCS) form for an isotropic $s$-wave gap, but the low-$T$ pseudogap is closely fit to that of a $d_{x^2-y^2}$-wave gap. Note that the superconducting gap dissapears at or near to $T_c = 91$ K, but the pseudogap is still present at 104 K, well above $T_c$. Reprinted with permission from  Y. Zhong, Y. Wang, S. Han, Y.-R. Lv, W.-L. Wang, D. Zhang, H. Ding, Y.-M. Zhang, L. Wang, K. He, R. Zhong, H A. Schneeloch, G. D. Gu, C.-L. Song, X.-C. Ma, and Q. K. Xue, Nodeless pairing in superconducting copper-oxide monolayer films on Bi$_2$Sr$_2$CaCu$_2$O$_{8+\delta}$, Sci. Bull. {\bf 61}, 1239 (2016). \cite{Zhong}}}
\end{figure}
\begin{figure}
\center{\includegraphics[width=0.3\textwidth]{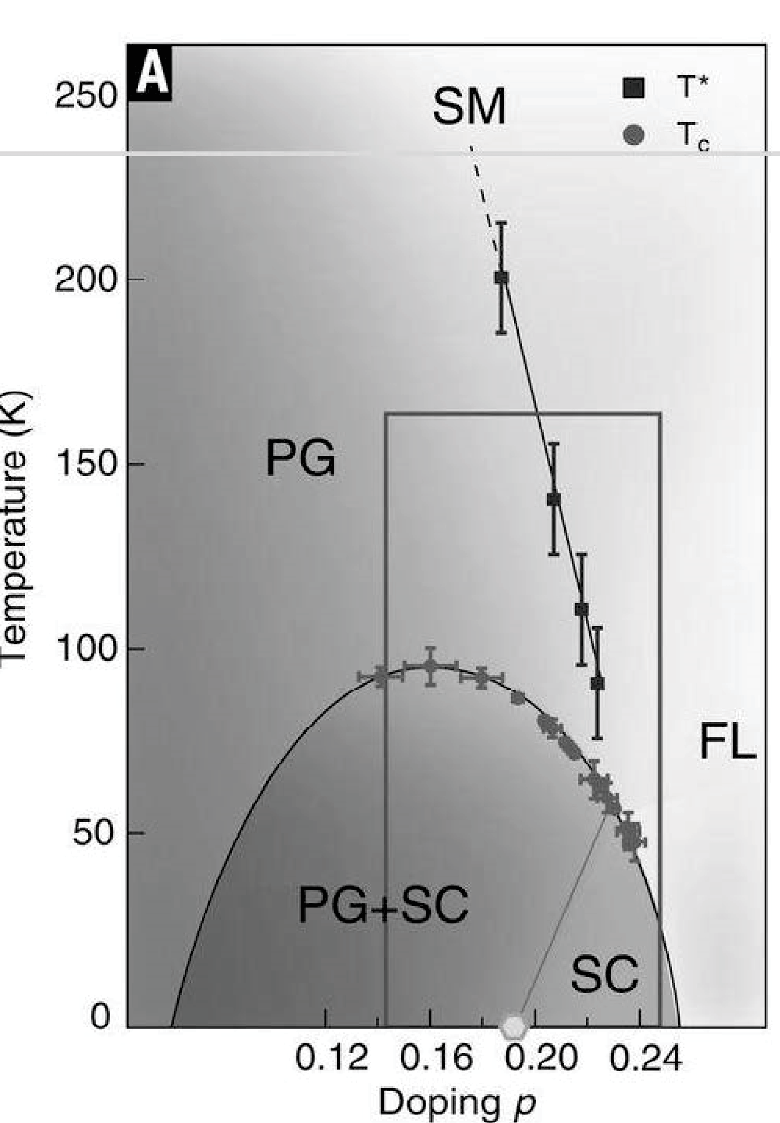}
\caption{Phase diagram of Bi2212 from the Stanford ARPES group.  $T^*$ is the pseudogap (PG) onset temperature, ând the regions labelled SM, FL, and SC  are respectively a strange metal,  an ordinary Fermi liquid, and a superconductor.  Note the boundary between strange and ordinary is at or beyond the critical concentration of 0.19, depending upon the temperature $T$.  Reprinted with permission from Y. He, M. Hashimoto, D. Song, S.-D. Chen, J. He, I. M. Vishik, B. Moritz, D.-H. Lee, N. Nagaosa, J. Zaanen, T. P. Devereaux, Y. Yoshida, H. Eisaki, D. H. Lu, and Z.-X. Shen, Rapid change of superconductivity and electron-phonon coupling through critical doping in Bi-2212, Science {\bf 362},  62-65 (2018). \cite{ZXShen}}}
\end{figure}

 \begin{figure}
\center{\includegraphics[width=0.4\textwidth]{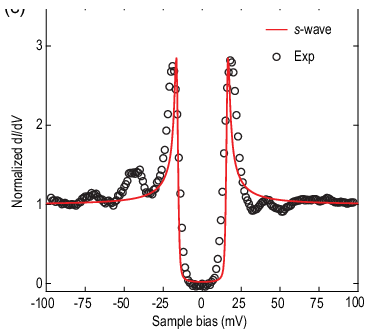}
\caption{The  density of states of the electron-doped cuprate Sr$_{1-x}$Nd$_x$CuO$_2$, which exhibits an $s$-wave superconducting gap and three optical phonon modes at energies above that of the superconducting gap, but no apparent quasilinear pseudogap contributions. Reprinted with permission from  J.-Q. Fan, X.-Q. Yu, F. J. Cheng, H. Wang, R. Wang, X. Ma X. Li, Q. Zhang, L. Gu, X. J. Zhou, J. Zhu, D. z, X.-P. Hu, D. Zhang, X.-C. Ma, Q.-K. Xue, and C.-L. Song, Direct observation of nodeless superconductivity and phonon modes in electron-doped copper oxide Sr$_{1-x}$Nd$_x$CuO$_2$, Nat. Sci. Rev. {\bf 9}, nwab225 (2022). doi: 10.1093/nsr/nwab225.\cite{electron-doped}}}
\end{figure}
\vskip20pt

Angle-resolved photoemission spectroscopy  (ARPES) experiments on Bi2212 have been abundant, and there is a general consensus that the pseudogap (PG) and superconducting (SC) regions of the phase diagram (depending upon the oxygen doping parameter $\delta$) are similar to that pictured in Fig. 5 \cite{ZXShen}.  Note that one requires a huge amount of oxygen annealing to get into the superconducting dome region that does not also contain the pseudogap.   Recently it was shown that  the electron-doped cuprate Sr$_{1-x}$Nd$_x$CuO$_2$ with one CuO$_2$ layer per unit cell, exhibits an $s$-wave superconducting gap and three optical phonon modes, but no CDW, consistent with the Hall $R_H(T)$ for 2$H$-TaS$_2$  shown in Fig. 1 \cite{electron-doped}. The density of states of Sr$_{1-x}$Nd$_x$CuO$_2$ is pictured in Fig. 6.

 \section{$c$-Axis twist experiments on Bi2212 and Bi2201}

  Several groups performed variations of the $c$-axis  twist experiment on Bi2212 \cite{Takano1,Takano2,Takano3,Takano4,Latyshev,Poccia,Zhu,Xue,JuHongLee,Zhao1}.   Aside from the the original $c$-axis twist experiment performed just below $T_c$ after annealing in excess oxygen \cite{Li}, the only recent experiment to do this was just published \cite{Xue}.  
 Those $c$-axis twist experiments on few-layer Bi2212 were performed both in the optimally-doped and overdoped regimes \cite{Xue}, which regions are indicated by the red  ``this work'' in Fig. 7.  In order to apply electrical leads in the overdoped regime, they had to temporarily stop the excess oxygen flow, which broadened the resistive transitions.  The Harvard group prepared an overdoped sample \cite{Zhao1}, but did not present any data for it \cite{Kim}.  From the ARPES and twist experiments, it is apparent that the overdoped regime is the crucial regime to study, as it contains only one order parameter, consistent with the region labeled SC in the Stanford ARPES phase diagram shown in Fig. 5.  Note that the region labelled CO (charge-ordered) in Fig. 7 correctly implies some form of a CCDW.

 \begin{figure}
\center{\includegraphics[width=0.49\textwidth]{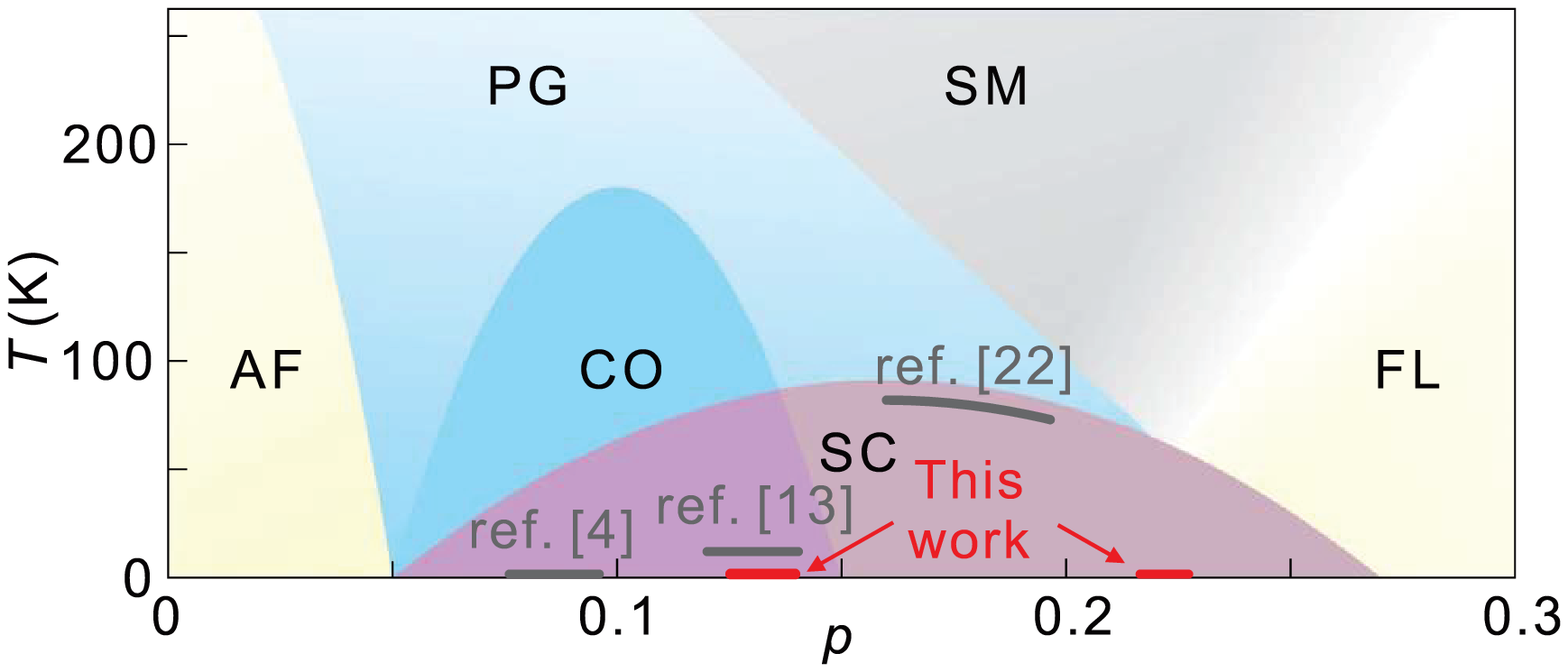}
\caption{Phase diagram of Bi2212 from the Tsinghua group and the highlighted  optimally-doped and overdoped regions in which their 45$^{\circ}$ $c$-axis twist experiments were performed. The data from the two regions labelled ``this work'' in red are from the Tsinghua Xue group \cite{Xue}, the data from the region labelled ``''[Ref 22]'' are from the original Li {\it et al.} $c$-axis twist experiment \cite{Li}, the data from the  region labelled ``[Ref 4]'' are also from the Tsinghua Xue group \cite{Zhu}, and the data from the region labelled ``[Ref 13]'' are from the Harvard Kim group \cite{Zhao1} . Reprinted with permission from Y. Zhu, H. Wang, Z. Wang, S. Hu, G. Gu, J. Zhu, D. Zhang, and Q.-K. Xue, Persistent Josephson tunneling between Bi$_2$Sr$_2$CaCu$_2$O$_{8+\delta}$ flakes twisted by 45$^{\circ}$ across the superconducting dome, Phys. Rev. B {\bf 108}, 174508 (2023). \cite{Xue}}}
\end{figure}

However, the experiments appear to be easier to perform on Bi2201 than on Bi2212.   Very recently, such $c$-axis twist experiments were performed on the monolayer Bi$_2$Sr$_{2-x}$La$_x$CuO$_{6+y}$ (Bi2201) \cite{Tsinghuapublished}.  They  focussed upon twist angles near to $45^{\circ}$, and their results are presented in Fig. 8.      Although they could not say that 100\% of the sample was consistent with $s$-wave superconductivity, they could infer from their data that a substantial amount of it had to have $s$-wave symmetry.

  \begin{figure}
 \center{\includegraphics[width=0.49\textwidth]{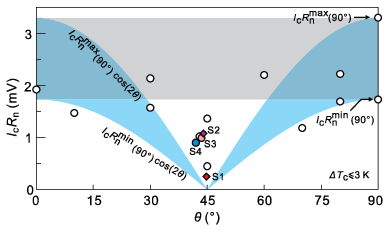}
\caption{Low-temperature data for $45^{\circ}$ $c$-axis twist junctions of ultra-thin films of Bi2201, along with some low-temperature twist data at different twist angles.  Reprinted with permission from H. Wang, Y. Zhu, Z. Bai, Z. Wang, S. Hu, H.-Y. Xie, X. Hu, J. Cui, M. Huong, J. Chen, Y. Ding, L. Zhao, X. Li, Q. Zhang, L. Gu, X. J. Zhou, J. Zhu, D. Zhang, and Q.-K. Xue, Prominent Josephson tunneling between twisted single copper oxide planes of Bi$_2$Sr$_{2-x}$La$_x$CuO$_{6+y}$, Nat. Commun. (2023) 14:501 doi.org/10.1038/s41467.023-405254.\cite{Tsinghuapublished}}}
\end{figure}

  \section{Distinguishing the superconducting gap from the pseudogap in Bi2201 and Bi2212}
  \begin{figure}
\center{\includegraphics[width=0.49\textwidth]{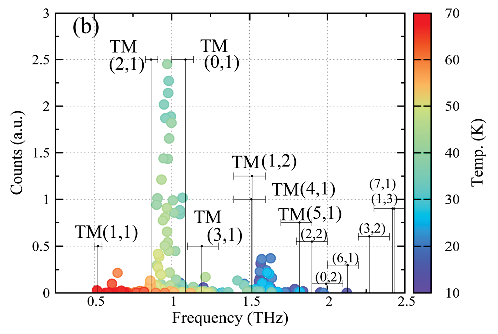}
\caption{Frequency dependence of the emission from a standalone Bi2212 disk mesa \cite{Kashiwagi1}.  Reprinted with permission from T. Kashiwagi, K. Sakamoto, H. Kubo, Y. Shibano, T. Enomoto, T. Kitamura, K. Asunuma, T. Yasui, C. Watanabe, K. Nakade, Y. Saiwai, T. Katsuragawa, M. Tsujimoto, R. Yoshizaki, T. Yamamoto, H. Minami, R. A. Klemm, and K. Kadowaki, A high-$T_c$ intrinsic Josephson junction emitter tunable from 0.5 to 2.4 terahertz, Appl. Phys. Lett. {\bf 107}, 082601 (2015) \copyright 2015 AIP Publishing LLC. }}
\end{figure}

\begin{figure}
\center{\includegraphics[width=0.49\textwidth]{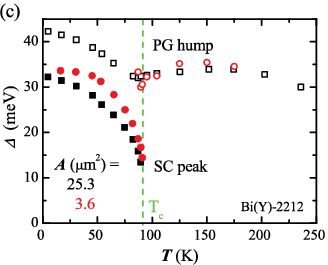}
\caption{Comparison of the temperature dependencies of the pseudogap and the superconducting gap in a magnetic field for two different slices of the same sample of Bi$_{2+x}$Sr$_{2-y}$CuO$_{6+\delta}$ (Bi2201).  Reprinted with permission from Th. Jacobs, S. O. Katterwe, H. Motzkau, A. Rydh, A. Maljuk, T. Helm, C. Putzke, E. Kampert, M. V. Kartsovnik, and V. M. Krasnov, Electron-tunneling measurements of low-$T_c$ single-layer Bi$_{2+x}$Sr$_{2-y}$CuO$_{6+\delta}$:  Evidence for a scaling disparity between superconducting and pseudogap states,  Phys. Rev. B {\bf 86}, 214506 (2012).\cite{KrasnovBi2201}}}
\end{figure}

  \begin{figure}
\center{\includegraphics[width=0.49\textwidth]{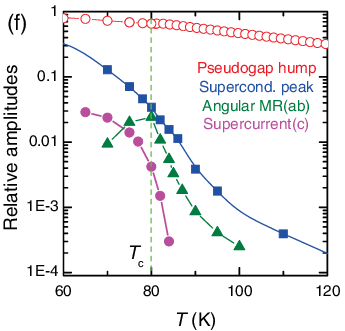}
\caption{Plotted are the logarithms of the pseudogap hump (upper open red circles), the superconducting peak (solid blue squares), the 2D cusp in the planar magnetoresistance (solid green triangles), and the Josephson $c$-axis supercurrent (solid purple circles) of Bi2212. Reprinted with permission from Th. Jacobs, S. O. Katterwe, and V. M. Krasnov, Superconducting correlations above $T_c$ in the pseudogap state of Bi$_2$Sr$_2$CaCu$_2$O$_{8+\delta}$ cuprates revealed by angular-dependent magnetotunneling, Phys. Rev. B {\bf 94}, 220501(R), (2016). \cite{KrasnovBi2212}}}
\end{figure}
 
\begin{figure}
\center{\includegraphics[width=0.49\textwidth]{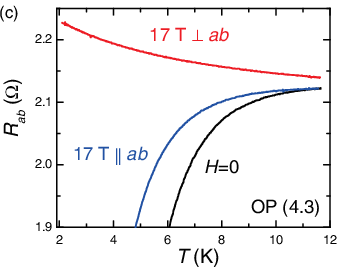}
\caption{Plotted are the resistance $R_{ab}(T)$ in the $ab$ plane of Bi2201 with  $T_c\approx 4.3$K\ for $H=0$ and $H=17$ T parallel and perpendicular to the layers. Reprinted with permission from S. O. Katterwe, Th. Jacobs, A. Maljuk, and V. M. Krasnov, Low anisotropy of the  upper critical field in a strongly anisotropic layered cuprate Bi$_{2.15}$Sr$_{1.9}$CuO$_{6+\delta}$:  Evidence for a paramagnetically limited superconductivity, Phys. Rev. B {\bf 89}, 214516 (2014).\cite{KrasnovBi2201Hc2}}}
\end{figure}

Since the superconducting gap in Bi2212 is entirely consistent with an isotropic BCS gap, to what should we attribute the pseudogap also observed in Bi2212?
An obvious possibility is that the pseudogap arises from a charge-density wave (CDW) \cite{WilsonDiSalvoMahajan}. As mentioned in the introductrion,
it has long been known that the transition metal dichalcogenides are also quasi-two-dimensional metals, some of which exhibit both superconductivity and CDWs \cite{WilsonDiSalvoMahajan,GDKG,Thompson,book,Tonjes,TMDsuperconductors}.  In particular, the $2H$ forms such as $2H$-TaS$_2$ and  $2H$-NbSe$_2$  were known to be both superconducting and to exhibit  CDWs.  Pristine $2H$-TaS$_2$ was known to be superconducting below $T_c\approx0.6$K.  Upon intercalation with pyridine, its $T_c$ increased dramatically to about 3.5 K \cite{GDKG}, which was later found to be due to suppression of the CDW by the intercalation process, which separated the individual layers by about 0.6 nm. Hence, it is apparent that the CDW is stabilized by interlayer interactions. Many transition metal dichalcogenides were subsequently found to exhibit CDWs, which were either incommensurate with the crystal lattice (ICDWs) or commensurate with it (CCDWs) \cite{WilsonDiSalvoMahajan,book}. Also, in layered compounds, the $T$ region of substantial superconducting fluctuations extends only up to about 3$T_c$ \cite{book,KLB},  which is strongly violated by the heavily underdoped ``pseudogap'' regions of the cuprates.

 Unlike the STM results on electron-doped Sr$_{1-x}$Nd$_x$CuO$_2$ shown in Fig. 6,  the problem with STM experiments on few-layer Bi2212 systems at low $T$ is that one has to be sure that the samples have oxygen stoichiometries that are beyond the  critical oxygen concentration of 0.19-0.22, depending upon $T$, as indicated in Fig. 5, \cite{ZXShen} below (or beyond) which the normal state properties of Bi2212 are very similar to the normal state properties of a conventional superconductor, so that the nodal CDW (or pseudogap) that greatly complicates the superconducting analysis is completely absent.  Since annealing at low $T$ in a heavy oxygen concentration is technically difficult \cite{Xue}, it would be better still to perform such few-layer,  low-temperature $c$-axis twist experiments on the single-layer Bi2201, in which the evidence for a nodal CDW has been weaker.  The Tsinghua group recently performed $c$-axis twist experiments on  Bi2201 \cite{Tsinghuapublished}, and their results are presented in Fig. 8. They concluded that the superconducting order parameter contains at least  a substantial amount of $s$-wave symmetry.
 
 \section{THz emission from cylindrical disk mesas of Bi2212}
There have been many experiments demonstrating THz emission from thin mesas of Bi2212 \cite{KK,Kashiwagi1,Kashiwagi2,Delfanazari,Tsujimoto,Benseman}  Most experiments were performed using single square or rectangular mesas, or using several evenly-spaced rectangular mesas  parallel to one another\cite{Delfanazari,Benseman}.
 In Fig. 9, the low-frequency region of the terahertz emission from cylindrical disk mesas of Bi2212 is shown \cite{Kashiwagi1}.  First, it should be noted that the vertical lines correspond to the predicted emission frequencies without the semiconducting substrate factors, which typically reduce the emission frequencies by about 3\% \cite{Balanis}.  This implies that the largest emission peak is identified as arising from the TM(0,1) mode \cite{Rain}.  A table of the predicted frequencies appropriate for the actual cylindrical disk was given previously \cite{Rain}.  More important is the fact that the emission linewidths are very narrow.  For an $s$-wave superconductor, the width of the emission  at the frequency $f$ would correspond to the width of the theoretical right BCS peak in Eq. (1) at the appropriate measurement energy $E_m$, or experimentally to the right peak in Fig. 3 or to that of the right peak in the black data taken at 4.2 K in the left panel of Fig. 4, also at the appropriate measurement energy $E_m$.
 
  For a $d_{x^2-y^2}$-wave superconductor, the theoretical linewidths could be calculated by numerically integrating the density of states given by Eq. (3) up to  $E_m$, the energy of the measurement, or experimentally to the width of the right black curve taken at 4.2 K in the right panel of Fig. 4 at $E_m$. It is easy to see that the narrow line widths in  the data of Fig. 9  provide very strong evidence that the superconducting state of Bi2212 has $s$-wave symmetry.  Therefore, if there is some component to the superconducting state of Bi2212 that has $d_{x^2-y^2}$-wave symmetry, it does not emit photons when a voltage is applied across the intrinsic Josephson junctions in Bi2212.  Moreover, since Bi2212 is a hole-doped layered superconductor, we may safely assume as in Fig. 1 that it is susceptible to CDWs, and in particular, to a CCDW, which we assume to arise on the oxygen sites in the CuO$_2$ layers.

 In Figs. 10 and 11, comparisons of the superconducting and pseudogap components of the order parameters in Bi2201 and Bi2212 as presented by the Krasnov group in Stockholm are shown \cite{KrasnovBi2201,KrasnovBi2212}.  As is evident in both figures, the superconducting components have standard mean-field behavior, disappearing at $T_c$, but the pseudogaps are present well above $T_c$. In the case of Bi2201, a sample with a $T_c$ of 4 K was studied, and it could be driven normal in a strong magnetic field, but the pseudogap was unaffected by such magnetic field strengths, as shown in Fig. 12 \cite{KrasnovBi2201Hc2}.  Hence, we assume that the pseudogaps in Bi2201, Bi2212, and also in Bi2223 are not superconducting phases, but are instead some forms of insulating states. The most likely insulating states are assumed to be CDWs.  Here, we assume that the CDWs that can be somewhat useful are the CCDWs.  We note that in the transition metal dichalcogenides, the relevant atoms they all have in common are the chalcogens, S, Se, and Te.  Thus it seems obvious that in the cuprates, the O atoms in the CuO$_2$ layers are the sites on on which the CCDWs form, as O is in the same 16$^{\rm th}$ column of the periodic table of the elements as are the chalcogens.
\section{The Model} 

In the absence of a CDW, the effective charge on each O site is $\langle q\rangle$, the spatial average of the O charges.  However, when a CDW begins to form, the charges on the individual O sites differ slightly from this average value.  This difference can be a small difference continuous from 0 in $T-T_{CDW}$  at $T_{CDW}$ for a 2nd order phase transition, as for a transition from a metallic to an ICDW state by Fermi surface nesting, or a  more substantial value discontinuous in $T-T_{CDW}$  at $T_{CCDW}$ for a 1st order transition, as in $2H$-TaS$_2$ at 75 K.

Our model for the  CCDWs in Bi2201, Bi2212 or Bi2223 is very simple.  The oxygen sites in a single  CuO$_2$ layer of Bi2201 are sketched in Fig. 13.   In this figure, effective charges $\pm q=\pm\delta e$ are assigned to neighboring O sites in orange and blue, respectively, and the central point with charge 0 is indicated in green. In Fig. 14, a sketch of a $d_{x^2-y^2}$-wave function is shown.  Note that the figure is rotated by 45$^{\circ}$ about the $c$-axis normal to the CuO$_2$ plane from that of Fig. 13.   The oxygen sites in the  double CuO$_2$ layer of Bi2212 are sketched in Fig. 15. The oxygen sites in the  triple CuO$_2$ layer of Bi2223 are sketched in Fig. 16.
 \begin{figure}\vskip10pt\center{\includegraphics[width=0.35\textwidth]{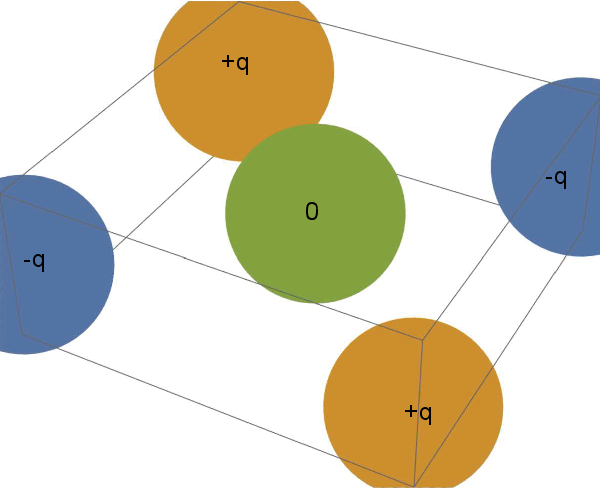}
\caption{Sketch of the symmetry of the CDW order parameter in a single CuO$_2$ layer.  The orange and blue dots correspond respectively to $\pm q$ excess charges on the oxygen sites, where where $q=\delta e$ and  $e$ is the electronic charge magnitude.  The central green dot represents the center of symmetry of  the CDW in that layer.  It has zero total charge in the ground state CDW wave function, which has $d_{x^2-y^2}$-wave symmetry that is rotated by $\pm\pi/4$ about the central axis (through the central green dot of charge 0) normal to the plane. The black lines are guides to the eye.}}
\end{figure}
\vskip10pt   
 \begin{figure}\vskip10pt\center{\includegraphics[width=0.25\textwidth]{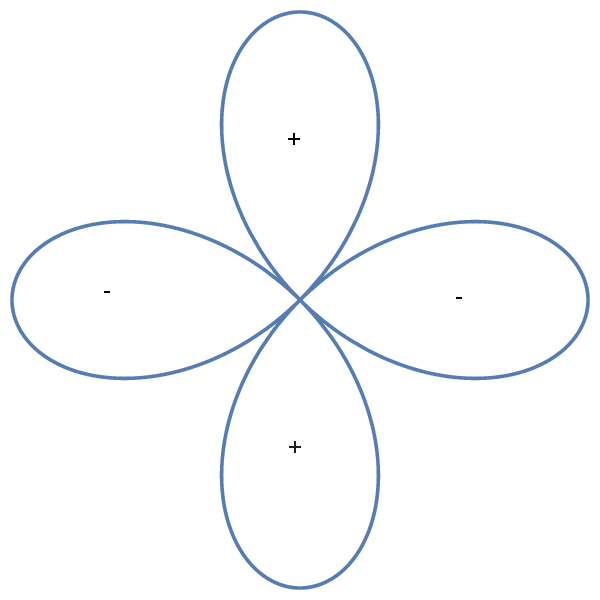}
\caption{Sketch of the symmetry of a ``conventional'' real-space  $d_{x^2-y^2}$-wave order parameter given by $\Delta(\theta)=\Delta_0\cos(2\theta)$, where $\theta=0$ corresponds to the $y$-axis.}}
\end{figure}
\vskip10pt   

\vskip5pt
\begin{figure}\vskip10pt\center{\includegraphics[width=0.45\textwidth]{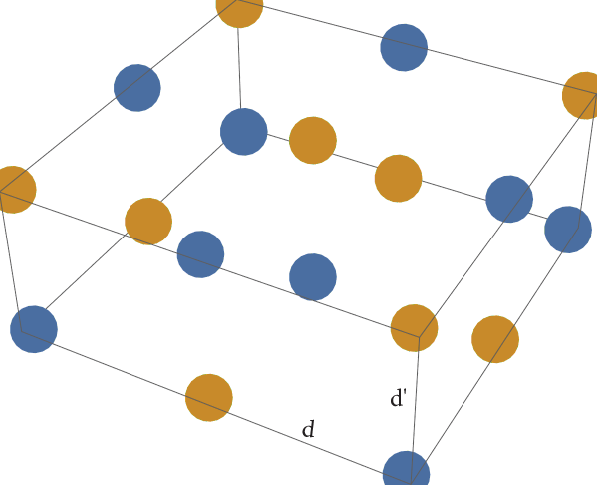}
\caption{Sketch of the ground state of the oxygen sites in a small section of a CuO$_2$ double layer.  The orange and blue dots correspond respectively to $\pm\delta e$ excess charges in the ground state, where $e$ is the electronic charge magnitude.  The black lines are guides to the eye.}}
\end{figure}
\vskip10pt

\begin{figure}\vskip10pt\center{\includegraphics[width=0.45\textwidth]{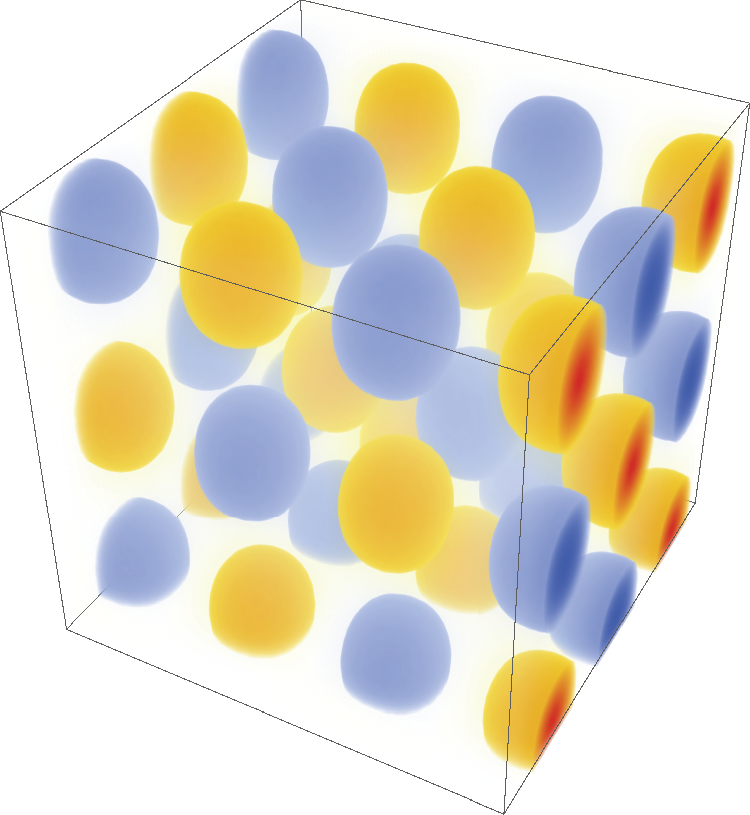}
\caption{Sketch of the ground state of the oxygen sites in a small section of a CuO$_2$ triple layer.  The orange and blue spheres correspond respectively to $\pm\delta e$ excess charges in the ground state, where $e$ is the electronic charge magnitude.  The black lines are guides to the eye.}}
\end{figure}
\vskip10pt

The effective Hamiltonian for the commensurate CDWs in Bi2201, Bi2212, and Bi2223 may be written  as an Ising model of  a rectangular prism of $LMN$ charges $\delta e\sigma_{i,j,k}$, each $\sigma_{i,j,k}$ with values $\pm1$, which may be written as
\begin{eqnarray}
H&=&H_{J}+ H_{J'},\\
H_J&=&-J\sum_{i=1}^{L-1}\sum_{j=1}^{M}\sum_{k=1}^{N}\sigma_{i,j,k}\sigma_{i+1,j,k}\nonumber\\
& &-J\sum_{i=1}^{L}\sum_{j=1}^{M-1}\sum_{k=1}^{N}\sigma_{i,j,k}\sigma_{i,j+1,k}\label{HJ}\\
H_{J'}& =&-J'\sum_{i=1}^{L}\sum_{j=1}^{M}\sum_{k=1}^{N-1}\sigma_{i,j,k}\sigma_{i,j,k+1},\label{HJ'}
\end{eqnarray}
where  $H_J$ and $H_{J'}$ are the intralayer and interlayer Hamiltonians,  and
\begin{eqnarray}
J&=&\frac{k(\delta e)^2}{d},\nonumber\\
J'&=&\frac{k(\delta e)^2}{d'},\label{Js}
\end{eqnarray}
 where $k$ is the Coulomb force constant in SI units.  Note that the interlayer oxygen-oxygen distance $d'$ is less than the near-neighbor intralayer oxygen distance $d$ in Bi2212 and Bi2223, so that  $J' > J >0$ \cite{book}.

In ths model,  $N=1$ and $J'=0$ for Bi2201, and $J'\ne0$ and respectively $N=2$ for Bi2212 and  $N=3$ for Bi2223.   The overall size of the general sublattice is therefore $L\times M\times N$.  Of course, the complete model has $L, M\rightarrow\infty$, which has only been solved exactly for the single layer \cite{Onsager}. 
 Note that the minus sign in each term of $H$ in Eq. (4)  accounts for the deviations $\delta e$ and $-\delta e$ on adjacent O sites from the overall average charge  $\langle q\rangle$ on each O site, which deviations develop due to  the Coulomb interaction of the site elements of the CCDW.  In addition,  the presence of the $N=2, 3$ layers is very important in stabilizing the CCDW formation, as shown in detail in the following.  Note that the Coulomb interaction favors opposite effective charge deviations from $\langle q\rangle$  on both intralayer and interlayer neighboring sites, which are incorporated into the overall negtive signs of all three terms in $H$.
 
In order to investigate the properties of this simple model, we first calculate the partition function $Z$.  From elementary statistical mechanics \cite{Shankar}, we have for an $(LMN)$ array of Ising spins $\sigma_{i,j,k}$,
\begin{eqnarray}
Z_{LMN}(K,K')&=&\sum_{\sigma_{i,j,k}=\pm1}\exp\Bigl(-H/(k_BT)\Bigr)
\end{eqnarray}
where  where $k_B$ is Boltzmann's constant and  $H$ is given by Eqs. (4) - (6).  We then define
\begin{eqnarray}
 K&=&J/(k_BT), \nonumber\\
 K'&=&J'/(k_BT)\label{KK'}.
 \end{eqnarray}
 For the single layer Bi2201 compound, we have
\begin{eqnarray}
Z_{LM1}(K)&=&\sum_{\sigma_{i,j}=\pm1}\exp\Biggl[K\Biggl(\sum_{i=1}^{L-1}\sum_{j=1}^{M}\sigma_{i,j}\sigma_{i+1,j}\nonumber\\
& &+\sum_{i=1}^{L}\sum_{j=1}^{M-1}\sigma_{i,j}\sigma_{i,j+1}\Biggr)\Biggr],
\end{eqnarray} 
and for general $N\ge2$, we have
 \begin{eqnarray}
Z_{LMN}(K,K')&=&\sum_{\sigma_{i,j,k}=\pm1}\exp\Biggl[K\sum_{k=1}^{N}\Biggl(\sum_{i=1}^{L-1}\sum_{j=1}^{M}\sigma_{i,j,k}\sigma_{i+1,j,k}\nonumber\\
& &+\sum_{i=1}^{L}\sum_{j=1}^{M-1}\sigma_{i,j,k}\sigma_{i,j+1,k}\Biggr)\nonumber\\
& &+K'\Biggl(\sum_{i=1}^{L}\sum_{j=1}^{M}\sum_{k=1}^{N-1}\sigma_{i,j,k}\sigma_{i,j,k+1}\Biggr)\Biggr].
\end{eqnarray}
Since we have already incorporated the opposite sign charges on neighboring O sites into the model, the largest interlayer contributions for $K'\rightarrow\infty$, which occurs as $T\rightarrow0$,
 occur for
\begin{eqnarray}
\sigma_{i,j,1}&=&\sigma_{i,j,2}=\ldots=\sigma_{i,j,N},
\end{eqnarray}
so that the interlayer factor is dominated by the term $\exp[K'LM(N-1)]$.
Similarly, the largest intralayer terms for $T\rightarrow0$ also occur for $K\rightarrow\infty$, which lead to 
\begin{eqnarray}
\sigma_{i,j,k}&=&\sigma_{i,j+1,k}=\sigma_{i+1,j,k},
\end{eqnarray}
so that the dominant intralayer part of the partition function as $T\rightarrow0$ becomes $\exp[KNL(M-1)+KNM(L-1)]$.  Hence, in the $T\rightarrow 0$ limit, we have the general result
\begin{eqnarray}
Z_{LMN}(K,K')&\rightarrow&\exp[KNL(M-1)+KNM(L-1)]\nonumber\\
& &\times\exp[K'LM(N-1)].
\end{eqnarray}
Details of a significant number of portions of this partition function for $N=1,2,3$ in the $T\rightarrow0$ limit are presented in the Supplementary Information.

R. A. K.  acknowledges discussions with  Thomas J. Bullard, Shane A. Cybart,  Steven A. Kivelson, Reinhold Kleiner, and  Ding Zhang, and 
 was partially supported by the U. S. Air Force Office of Scientific Research (AFOSR) LRIR \#18RQCOR100, and the AFRL/SFFP Summer Faculty Program provided by AFRL/RQ at WPAFB.


\begin{thebibliography}{99}
\bibitem{book} R. A.  Klemm, Layered Superconductors, Vol. I (Oxford University Press, Oxford, UK, 2012).
\bibitem{Klemm1} R. A. Klemm, Striking similarities between the pseudogap phenomena in cuprates and in layered organic and dichalcogenide superconductors, Physica C {\bf 341.348}, 839.842 (2000).  doi:10.1016/S0921-4534(00)00708-5
\bibitem{TMDsuperconductors} R.  A. Klemm, Pristine and intercalated transition metal dichalcogenide superconductors, Physica C {\bf 514}, 86 (2015).
\bibitem{WilsonDiSalvoMahajan} J. A. Wilson, F. J. DiSalvo, and S. Mahajan,  Charge-density waves and superlattices in the metallic transition metal dichalcogenides, Adv. Phys. {\bf 24} (2), 117-201 (1975).  doi. 10.1080/00018737500101391.
\bibitem{Frindt} J. P. Tidman, O. Singh, D.E. Curzon, and R. F. Frindt, The phase transition in $2H$-TaS$_2$ at 75 K, Phil. Mag. {\bf 30}, 1191-1194 (1974).
\bibitem{Tonjes} W. C. Tonjes, V. A. Greanya, R. Liu, C. G. Olson, and P. Molinié, Charge-density-wave mechanism in the 2$H$-NbSe$_2$ family:  Angle-resolved photoemission studies, Phys. Rev. B {\bf 63}, 235101 (2001).
\bibitem{GDKG} F. R. Gamble, F. J. DiSalvo, R. A. Klemm, and T. H. Geballe, Superconductivity in layered structure organometallic crystals, Science {\bf 168}, 568 (1970).
\bibitem{Thompson} A. H. Thompson, F. R. Gamble, and R. F. Koehler, Effects of intercalation on electron transport in tantalum disulfide, Phys. Rev. B {\bf 5}(8), 2811 (1972).
\bibitem{Klemm3} R. A. Klemm, Fluctuation-induced conductivity in layered superconductors, J. Low Temp. Phys. {\bf 16}, 381 (1972).
\bibitem{KLB} R. A. Klemm, M. R. Beasley, and A. Luther, Fluctuation-induced diamagnetism un dirty three-dimensional, two-dimensional, and layered superconductors, Phys. Rev. B {\bf 8}, 5072 (1973).
\bibitem{VanHarlingen} D. A. Wollman, D. J. Van Harlingen, W. C. Lee, D. M. Ginsberg, and A. J. Leggett, Experimental determination of the superconducting pairing state in YBCO from the phase coherence of YBCO-Pb dc SQUIDs, Phys. Rev. Lett. {\bf 71}, 2134 (1993).
\bibitem{TsueiKirtley1} C. C. Tsuei, J. A. Kirtley, C. C. Chi, S. S. Yu-Jahnes, A. Gupta, T. Shaw, J. Z. Sun, and M. B. Ketchen, Pairing symmetry and flux quantization in a tricrystal superconducting ring of YBa$_2$Cu$_3$O$_{7-\delta}$, Phys. Rev. Lett. {\bf 73}, 593 (1994).
\bibitem{TsueiKirtley2} C. C. Tsuei and J. A. Kirtley, Pairing symmetry in cuprate superconductors, Rev. Mod. Phys. {\bf 72}, 969-1016 (2000).
\bibitem{KRS1} R. A. Klemm, C. T. Rieck, and K. Scharnberg, Angular dependence of the Josephson critical current in $c$-axis twist junctions of high temperature superconductors, Phys. Rev. B {\bf 58} (2), 1051 (1998).
\bibitem{KARS} R. A. Klemm, G. Arnold, C. T. Rieck, and K. Scharnberg, Coherent vs. incoherent $c$-axis Josephson tunneling between layered superconductors,  Phys. Rev. B {\bf 58} (21), 14203 (1998).
\bibitem{Li} Q. Li, Y. N. Tsay, M. Suenaga, R. A. Klemm, G. D. Gu, and Y. Koshizuka,  Bi$_2$Sr$_2$CaCu$_2$O$_{8+\delta}$  Bicrystal $c$-axis twist Josephson junctions:  A new phase-sensitive test of order parameter symmetry, Phys. Rev. Lett. {\bf 83} (20), 4160  (1999).
\bibitem{KRS2} R. A. Klemm, C. T. Rieck, and K. Scharnberg,  Order parameter symmetries in high temperature superconductors, Phys. Rev. B {\bf 61} (9), 5913 (2000).
\bibitem{BKS} A. Bille, R. A. Klemm,  and K. Scharnberg, Models of $c$-axis twist Josephson junctions,  Phys. Rev. B {\bf 64} (17), 174507 (2001).
\bibitem{Takano1} Y. Takano, T. Hatano, A. Ishii, A. Fukuyo, Y. Sato, S. Arisawa, and K. Togano, Fabrication of Bi2212 cross-whiskers junction, Physica C {\bf 362}, 261-264 (2001).
\bibitem{Takano2} Y. Takano, T. Hatano, A. Fukuyo, A. Ishii, M. Ohmori, S. Arisawa, K. Togano, and M. Tachiki, $d$-like Symmetry of the order parameter and intrinsic Josephson effects in Bi$_2$Sr$_2$CaCu$_2$O$_{8+\delta}$ cross-whisker junctions, Phys. Rev. B {\bf 65}, 140513 (2002).
\bibitem{Takano3} Y. Takano, T. Hatano, M. Ohmori, S. Kawakami, A. Ishii, S. Arisawa, S.-J. Kim, T. Yamashita, K. Togano, and M. Tachiki, Cross-whisker intrinsic Josephson junction as a probe of the superconducting order parameter, J. Low Temp. Phys. {\bf 131}, 533-537 (2003).
\bibitem{Takano4} Y. Takano, T. Hatano, A. Fukuyo, M. Ohmori, P. Ahmet,  T. Naruke, K. Nakajima, T. Chikyow, A. Ishii, S. Arisawa, K. Togano, and M. Tachiki, Angular dependence of critical current and intrinsic Josephson effects in Bi-2212 cross-whisker junctions, Sing. J. Phys. {\bf 18}, 67 (2002).
\bibitem{Klemm2} R. A.  Klemm, Theory of   Bi$_2$Sr$_2$CaCu$_2$O$_{8+\delta}$ cross-whisker Josephson junctions,  Phys. Rev. B {\bf 67}, 174509 (2002).
\bibitem{Latyshev} Yu. I. Latyshev, A. P. Orlov, A. M. Nikitina, P. Monceau, and R. A.  Klemm, The $c$-axis transport in naturally-grown Bi$_2$Sr$_2$CaCu$_2$O$_{8+\delta}$ cross-whisker junctions, Phys. Rev. B {\bf 70}, 094517 (2004).
\bibitem{KlemmPhilMag}  R. A. Klemm, The phase-sensitive $c$-axis twist experiments on Bi$_2$Sr$_2$CaCu$_2$O$_{8+\delta}$ and their implications, Phil. Mag. {\bf 85}, 801-853 (2005).
\bibitem{KivelsonSpivak} S. A. Kivelson and B. Spivak, Macroscopic character of composite high-temperature superconducting wires, Phys. Rev. B {\bf 92}, 184502 (2015).
\bibitem{Kadowaki} B. W. Hoogenboom, K. Kadowaki, B. Revaz, and {\O}. Fischer, Homogeneous samples of Bi$_2$Sr$_2$CaCu$_2$O$_{8+\delta}$, Physica C {\bf 391}, 376-380 (2003). 
\bibitem{Zhong} Y. Zhong, Y. Wang, S. Han, Y.-R. Lv, W.-L. Wang, D. Zhang, H. Ding, Y.-M. Zhang, L. Wang, K. He, R. Zhong, H A. Schneeloch, G. D. Gu, C.-L. Song, X.-C. Ma, and Q. K. Xue, Nodeless pairing in superconducting copper-oxide monolayer films on Bi$_2$Sr$_2$CaCu$_2$O$_{8+\delta}$, Sci. Bull. {\bf 61}, 1239 (2016).
\bibitem{BCS}  J. Bardeen, L. N. Cooper, and J. R. Schrieffer, Theory of Superconductivity, Phys. Rev. {\bf 108}, 175 (1957). 
\bibitem{Mahan} G.D. Mahan, Condensed Matter in a Nutshell (Princeton University Press, 2011).
\bibitem{SK1} K. Scharnberg and R. A. Klemm, $P$-wave superconductors in magnetic fields, Phys. Rev. B {\bf 22}, 5233 (1980 ).
\bibitem{SK2}  K. Scharnberg and R. A. Klemm, Upper critical field in $p$-wave superconductors with broken symmetry, Phys. Rev. Lett. {\bf 54}, 2445 (1985).
\bibitem{ZXShen} Y. He, M. Hashimoto, D. Song, S.-D. Chen, J. He, I. M. Vishik, B. Moritz, D.-H. Lee, N. Nagaosa, J. Zaanen, T. P. Devereaux, Y. Yoshida, H. Eisaki, D. H. Lu, and Z.-X. Shen, Rapid change of superconductivity and electron-phonon coupling through critical doping in Bi-2212, Science {\bf 362},  62-65 (2018).
\bibitem{electron-doped} J.-Q. Fan, X.-Q. Yu, F. J. Cheng, H. Wang, R. Wang, X. Ma, X.-P. Hu, D. Zhang, X.-C. Ma, Q.-K. Xue, and C.-L. Song, Direct observation of nodeless superconductivity and phonon modes in electron-doped copper oxide Sr$_{1-x}$Nd$_x$CuO$_2$, Nat. Sci. Rev. {\bf 9}, nwab225 (2022). doi: 10.1093/nsr/nwab225.
\bibitem{Poccia} M. Markini, Y. Lee, T. Confalone, S. Shokrin, C. N. Sagaru, D. Wolf, G. Gu, K. Watanabe, T. Taniguchi, D. Montemurro, V. M. Vinokour, K. Nielsch, and N. Poccia, Twisted cuprate van der Waals heterostructures with controlled Josephson coupling, Mat. Today {\bf 67}, 106-112 (2023].
\bibitem{Zhu} Y. Zhu, M. Liao, Q. Zhang, H.-Y. Xie, F. Meng, Y. Liu, Z. Bai, S. Ji, J. Zhang, K. Jiang, R. Zhong, J. Schneeloch, G. Gu, L. Gu, X. Ma, D. Zhang, and Q.-K. Xue, Presence of $s$-wave pairing in Josephson junctions made of twisted ultrathin Bi$_2$Sr$_2$CaCu$_2$O$_{8+x}$ flakes, Phys. Rev. X {\bf 11}, 031011 (2021).
\bibitem{Xue} Y. Zhu, H. Wang, Z. Wang, S. Hu, G. Gu, J. Zhu, D. Zhang, and Q.-K. Xue, Persistent Josephson tunneling between Bi$_2$Sr$_2$CaCu$_2$O$_{8+\delta}$ flakes twisted by 45$^{\circ}$ across the superconducting dome, Phys. Rev. B {\bf 108}, 174508 (2023).
\bibitem{JuHongLee} J. Lee, W. Lee, G. Y. Kim, V. B. Choi, J. Park, S. Jang, G. Gu, S. Y. Choi, G. Y. Cho, G. H. Lee, and H. J. Lee, Twisted van der Waals Josephson junction based on a high-$T_c$ superconductor, Nano Lett. {\bf 21} (21), 10469-10477 (2021).
\bibitem{Zhao1} S. Y. F. Zhao, X. Cui, P. A. Volkov, H. Yoo, S.Lee, J. A. Gardener, A. J. Akey, R. Engelke, Y. Ronen, R. Zhong, G. Gu, S. Plugge, T. Tummuru, M. Franz, J. H. Pixley, N. Poccia,  and P. Kim, Time-reversal symmetry breaking superconductivity between twisted cuprate superconductors, Science                                               {\bf 382}, 6677, pp. 1422-1427 doi:10.1126/science.abl8371 (2023).
\bibitem{Kim} When the senior author of \cite{Zhao1} was asked by the corresponding author of this manuscript at the 13th International Conference on the Intrinsic Josephson Effect and High Temperature Superconductivity in Glasgow, Scotland, August 30-September 1, 2023 why he didn't show any data for their  overdoped sample, he replied that there was excess oxygen in its grain boundary.  We note that all overdoped samples studied to date have excess oxygen in their grain boundaries.
\bibitem{Tsinghuapublished} H. Wang, Y. Zhu, Z. Bai, Z. Wang, S. Hu, H.-Y. Xie, X. Hu, J. Cui, M. Huong, J. Chen, Y. Ding, L. Zhao, X. Li, Q. Zhang, L. Gu, X. J. Zhou, J. Zhu, D. Zhang, and Q.-K. Xue, Prominent Josephson tunneling between twisted single copper oxide planes of Bi$_2$Sr$_{2-x}$La$_x$CuO$_{6+y}$, Nat. Commun. (2023) 14:5201 doi.org/10.1038/s41467.023-405254.
\bibitem{KK} R. A. Klemm and K. Kadowaki, Output from a Josephson stimulated terahertz amplified radiation emitter, J. Phys.: Condensed Matter {\bf 22}, 3745701 (2010).
\bibitem{Kashiwagi1} T. Kashiwagi, K. Sakamoto, H. Kubo, Y. Shibano, T. Enomoto, T. Kitamura, K. Asanuma, T. Yasui, C. Watanabe, K. Nakade,  Y. Saiwai, T. Katsuragawa, M. Tsujimoto, R. Yoshizaki, T. Yamamoto,  H. Minami, R. A. Klemm, and K. Kadowaki,   A high-$T_c$ intrinsic Josephson emitter tunable from 0.5 to 2.4 terahertz, Appl. Phys. Lett. {\bf 107}, 082601 (2015). DOI: 10.1063/1.4929715.
\bibitem{Kashiwagi2} T. Kashiwagi, T. Yamamoto, H. Minami, M. Tsujimoto, R. Yoshizaki, K. Delfanazari, T. Kitamura, C. Watanabe, K. Nakade, T. Yasui, K. Asanuma, Y. Saiwai, Y. Shibano, T. Enomoto, H. Kubo, K. Sakamoto, T. Katsuragawa, B. Markovi{\'c}, J. Mirkovi{\'c}, R. A. Klemm, and K. Kadowaki, Efficient fabrication of intrinsic Josephson junction terahertz oscillators with greatly reduced self-heating effects, Phys. Rev. Applied {\bf 4}, 054018 (2015).  DOI: 10.1103/PhysRevApplied.4.054018.
\bibitem{Delfanazari} K. Delfanazari, R. A. Klemm, H. J. Joyce, D. A. Ritchie, and K. Kadowaki,  Integrated, portable, tunable, and coherent terahertz sources and sensitive detectors based on layered superconductors,  Proc. IEEE {\bf 108}, 721-734 (2020). DOI: 10.1109/JPROC.2019.2958810.
\bibitem{Tsujimoto} M. Tsuijimoto,   K. Yamaki, K. Deguchi, T. Yamamoto, T. Kashiwagi, H. Minami, M. Tachiki, K. Kadowaki, and R. A. Klemm,  Geometrical resonance conditions for THz radiation from the intrinsic Josephson junctions in Bi$_2$Sr$_2$CaCu$_2$O$_{8+\delta}$, Phys. Rev. Lett. {\bf 105}, 037005 (2010). 
\bibitem{Benseman} T. Benseman, K. E. Gray, A. E. Koshelev, W. K. Kwok, U. Welp, H. Minami, K. Kadowaki, and T. Yamamoto, Powerful terahertz emission from Bi$_2$Sr$_2$CaCu$_2$O$_{8+\delta}$ mesa arrays, Appl. Phys. Lett. {\bf 103}, 022602 (2013). doi: 10.1063/1.4813536.  
\bibitem{Balanis} C. A. Balanis, Antenna Theory, Analysis and Design, 3rd Ed., Hoboken, NJ, Wiley (2005).
\bibitem{Rain} J. R. Rain,   PeiYu Cai,  Alexander Baekey, Matthew A. Reinhard, Roman I. Vasquez, Andrew C. Silverman,  Christopher L. Cain, and Richard A. Klemm, Wavefunctions for high-symmetry, thin microstrip antennas, and two-dimesnional quantum boxes, Phys. Rev. A {\bf 104}, 062205 (2021).
 \bibitem{KrasnovBi2201} Th. Jacobs, S. O. Katterwe, H. Motzkau, A. Rydh, A. Maljuk, T. Helm, C. Putzke, E. Kampert, M. V. Kartsovnik, and V. M. Krasnov, Electron-tunneling measurements of low-$T_c$ single-layer Bi$_{2+x}$Sr$_{2-y}$CuO$_{6+\delta}$:  Evidence for a scaling disparity between superconducting and pseudogap states,  Phys. Rev. B {\bf 86}, 214506 (2012).
 \bibitem{KrasnovBi2212} Th. Jacobs, S. O. Katterwe, and V. M. Krasnov, Superconducting correlations above $T_c$ in the pseudogap state of Bi$_2$Sr$_2$CaCu$_2$O$_{8+\delta}$ cuprates revealed by angular-dependent magnetotunneling, Phys. Rev. B {\bf 94}, 220501(R), (2016).
 \bibitem{KrasnovBi2201Hc2} S. O. Katterwe, Th. Jacobs, A. Maljuk, and V. M. Krasnov, Low anisotropy of the  upper critical field in a strongly anisotropic layered cuprate Bi$_{2.15}$Sr$_{1.9}$CuO$_{6+\delta}$:  Evidence for a paramagnetically limited superconductivity, Phys. Rev. B {\bf 89}, 214516 (2014).
\bibitem{Onsager}  Lars Onsager, Crystal Statistics. I. A Two-Dimensional Model with an Order-Disorder Transition, Phys. Rev. {\bf 65}, 117 (1944).
\bibitem{Shankar} R. Shankar, Quantum Field Theory and Condensed Matter, An Introduction (Cambridge University Press, Cambridge, UK 2017).

\end{thebibliography}
 \end{document}


\title{Supplementary Material:  Model for the commensurate charge-density waves in under-hole-doped cuprate superconductors}

\author{Jacob Morrissey}
\affiliation{Department of Physics, University of Central Florida, Orlando, FL 32816-2385, USA}
\author{Timothy J. Haugan}
\affiliation{U. S. Air Force Research Laboratory, Wright-Patterson Air Force Base, Ohio 45433-7251, USA}
\author{Richard A. Klemm}
\email{richard.klemm@ucf.edu, corresponding author}
\affiliation{Department of Physics, University of Central Florida, Orlando, FL 32816-2385, USA}
\affiliation{U. S. Air Force Research Laboratory, Wright-Patterson Air Force Base, Ohio 45433-7251, USA}
\date{\today}

\begin{abstract}
A simple model of the  commensurate charge-density wave (CCDW) portion of the underdoped pseudogap regions of monolayer Bi$_2$Sr$_{2-x}$La$_x$CuO$_{6-x}$ (Bi2201), bilayer   Bi$_2$Sr$_2$CaCu$_2$O$_{8+\delta}$ (Bi2212), and trilayer Bi$_2$Sr$_2$Ca$_2$Cu$_3$O$_{10+\delta}$ (Bi2223) cuprate superconductors is presented and studied.  Some special cases of the model are studied in detail and presented here.
\end{abstract}

\pacs{} \vskip0pt
\maketitle

\title{Supplementary Material:   Model for the commensurate charge-density waves in under-hole-doped cuprate superconductors}

\author{Jacob Morrissey}
\affiliation{Department of Physics, University of Central Florida, Orlando, FL 32816-2385, USA}
\author{Timothy J. Haugan}
\affiliation{U. S. Air Force Research Laboratory, Wright-Patterson Air Force Base, Ohio 45433-7251, USA}
\author{Richard A. Klemm}
\email{richard.klemm@ucf.edu, corresponding author}
\affiliation{Department of Physics, University of Central Florida, Orlando, FL 32816-2385, USA}
\affiliation{U. S. Air Force Research Laboratory, Wright-Patterson Air Force Base, Ohio 45433-7251, USA}
\date{\today}

\section{$J'\rightarrow\infty$ Results for finite elements of sizes $L\times M\times N$ for $N = 2, 3$}

It is useful to first examine the one-dimensional array of charges on the two or three layers.  We first evaluate $Z_{112}$, which is the partition function for two neighboring sites on the two layers.    It is easly found to be
\begin{eqnarray}
Z_{112}(K')&=&\sum_{\sigma_{1,1},\sigma_{1,2}=\pm 1}\exp[-K'\sigma_{1,1}\sigma_{1,2}]\nonumber\\
&=&2(e^{K'}+e^{-K'})=4\cosh(K').\label{Z112}
\end{eqnarray}
Similarly, for the simplest three-layer case, we have
\begin{eqnarray}
Z_{113}(K')&=&\sum_{\sigma_{1,1},\sigma_{1,2},\sigma_{1,3}=\pm 1}\exp[-K'(\sigma_{1,1}\sigma_{1,2}+\sigma_{1,2}\sigma_{1,3})]\nonumber\\
&=&4+2(e^{2K'}+e^{-2K'})=4[1+\cosh(2K')]\nonumber\\
&=&8\cosh^2(K').\label{Z113}
\end{eqnarray}
Next, $Z_{212}$ is the partition function for two adjacent charges directly above two adjacent charges in the lower plane.  It is easily found to be
\begin{eqnarray}
Z_{212}(K,K')&=&\sum_{{\sigma_{1,1},\sigma_{1,2}=\pm 1}\atop{\sigma_{2,1},\sigma_{2,2}=\pm1}}\exp[-K'(\sigma_{1,1}\sigma_{1,2}+\sigma_{2,1}\sigma_{2,2})]\nonumber\\
& &\hskip30pt\times\exp[-K(\sigma_{1,1}\sigma_{2,1}+\sigma_{1,2}\sigma_{2,2})]\nonumber\\
&=&8[1+\cosh(2K)\cosh(2K')].\label{Z112}
\end{eqnarray}
For two adjacent charges in one layer,
\begin{eqnarray}
Z_{211}(K)&=&\sum_{\sigma_{1,1},\sigma_{2,1}=\pm1}\exp[-K\sigma_{1,1}\sigma_{2,1}]\nonumber\\
&=&4\cosh(K).
\end{eqnarray}
Then, we have for two adjacent charges in one layer adjacent to two adjacent charges in a second layer,
\begin{eqnarray}
\frac{Z_{212}(K,K')}{Z^2_{112}(K')}&=&\frac{8[1+\cosh(2K)\cosh(2K')]}{8[1+\cosh(2K')]}\nonumber\\
&\begin{array}{c}
\rightarrow\\ K'\rightarrow\pm\infty\end{array}&\cosh(2K)\\
&=&\frac{Z_{211}(2K)}{2^2}.
\end{eqnarray}
For the analogous three-layer case, we have
\begin{eqnarray}
\frac{Z_{213}(K,K')}{Z^2_{113}(K')}&=&\frac{8\cosh(4K')\cosh(K)(2\cosh(2K)-1)}{[4\cosh(2K')]^2}\hskip20pt\nonumber\\
&\begin{array}{c}
\rightarrow\\ K'\rightarrow\pm\infty\end{array}&\cosh(3K) =\frac{Z_{211}(3K)}{2^2}.
\end{eqnarray}

Cases for $Z_{L13}(K,K')$ for $L$ up to 10 are shown in the Supplementary Material.

We  note that we could have obtained the same result in the $T\rightarrow0$ limit, but as shown
 for $Z_{312}(K,K')/Z^3_{112}(K')$ in the Supplementary Material,
letting $T\rightarrow0$ leads to a formula that depends upon both $K$ and $K'$.    Hence, we can only take the limit $K'\rightarrow\infty$, which is equivalent to the $J'/J\rightarrow\infty$ limit.  In this limit, we obtain
\begin{eqnarray}
\frac{Z_{312}(K,K')}{Z^3_{112}(K')}&\begin{array}{c}
\rightarrow\\ K'\rightarrow\pm\infty\end{array}&\cosh^2(2K).
\end{eqnarray}
More generally, we have shown for $1\le L\le 10$ that
\begin{eqnarray}
\frac{Z_{L12}(K,K')}{Z^L_{112}(K')}&\begin{array}{c}
\rightarrow\\ K'\rightarrow\pm\infty\end{array}&\cosh^{(L-1)}(2K),
\end{eqnarray}
so that in this $K'\rightarrow\pm\infty$ limit, $Z_{212}(K,K')/Z^2_{112}(K')$ is an effective one-dimensional transfer matrix given by Eq. (9).

Analogously for the three-layer case, we have shown for $1 \le L\le 10$ that
\begin{eqnarray}
\frac{Z_{L13}(K,K')}{Z^L_{113}(K')}&\begin{array}{c}
\rightarrow\\ K'\rightarrow\pm\infty\end{array}&\cosh^{(L-1)}(3K),
\end{eqnarray}

We next considered the 3 cases of $L=M=2$.  It is easy to show that $Z_{221}(K)=4(\cosh(4K)+3)$.  Using Mathematica$^{\copyright}$ software, we found that $Z_{22N}$  for $N=2,3$  satisfies
\begin{eqnarray}
\frac{Z_{22N}(K,K')}{Z^N_{113}(K')}&\begin{array}{c}
\rightarrow\\ K'\rightarrow\pm\infty\end{array}&Z_{221}(NK).
\end{eqnarray}

Then, we considered the three cases of $L=3$ and $M=2$.  We found that $Z_{321}(K)=4[9\cosh(K)+6\cosh(3K)+\cosh(7K)]$, and that $Z_{32N}$ for $N=2,3$ both satisfy
\begin{eqnarray}
\frac{Z_{32N}(K,K')}{[Z_{113}(K')/2]^{N(N+1)}}&\begin{array}{c}
\rightarrow\\ K'\rightarrow\pm\infty\end{array}&Z_{321}(NK).
\end{eqnarray}

Then, we considered the three cases of $L=3$ and $M=3$.  We found that $Z_{331}(K)=4[36+48\cosh(2K)+23\cosh(4K)+16\cosh(6K)+4\cosh(8K)+\cosh(12K)]$, and that $Z_{33N}$ for $N=2,3$ both satisfy
\begin{eqnarray}
\frac{Z_{33N}(K,K')}{2\cosh(3^2(N-1)K')}&\begin{array}{c}
\rightarrow\\ K'\rightarrow\pm\infty\end{array}&Z_{331}(NK).
\end{eqnarray}

Then, we considered the three cases of $L=4$ and $M=2$.  We found that $Z_{421}=4[20+24\cosh(2K)+12\cosh(4K)+7\cosh(6K)+\cosh(10K)]$, and that $Z_{42N}$ for $N=2,3$ both satisfy
\begin{eqnarray}
\frac{Z_{42N}(K,K')}{2\cosh(2^{N+1}K')}&\begin{array}{c}
\rightarrow\\ K'\rightarrow\pm\infty\end{array}&Z_{421}(NK).
\end{eqnarray}

We have made use of Mathematica$^{\copyright}$ software to evaluate two-layer sections of sizes $L\times M\times2$ for $L,M\le4$ and for $(L,M)=(6,2), (5,3)$ and (5,2), and we found that  the general pattern at low $T$ satisfies
\begin{eqnarray}
\frac{Z_{LM2}(K,K')}{[Z_{112}(K')]^{LM}}&\begin{array}{c}
\rightarrow\\ K'\rightarrow\infty\end{array}&\frac{Z_{LM1}(2K)}{2^{LM}}.
\end{eqnarray}
There are two general forms for $Z_{LM1}(K)$.  These are
\begin{eqnarray}
Z_{LM1}^{(1)}(K)&=&\sum_{n=0}^{n_b/2}c_n\cosh(2nK)\label{even}
\end{eqnarray}
and
\begin{eqnarray}
Z_{LM1}^{(2)}(K)&=&\sum_{n=0}^{(n_b-1)/2}d_n\cosh[(2n+1)K]\label{odd},
\end{eqnarray}
where $n_b$ is the number of bonds between adjacent sites either within the same layer or on different layers, and $c_{n_b/2-1}=d_{(n_b-3)/2}=0$, except for the cases of $Z^{(1)}_{521}(K)$ and $Z^{(2)}_{411}(K)$ that we studied.  In all cases, we have for $K=0$, 
 \begin{eqnarray}
 Z_{LM1}^{(1)}(0)&=&\sum_{n=0}^{n_b/2}c_n \\
 &=&2^{LM}\\
  Z_{LM1}^{(2)}(0)&=&\sum_{n=0}^{(n_b-1)/2}d_n\\
  &=&2^{LM}.
 \end{eqnarray}
 A table of  single-layer connected sites $LM1$, double-layer connected sites $LM2$,   triple-layer connected sites $LM3$, and their respective number of neighboring site bonds $n_b$ that we studied is given in the Supplementary Material.

These simple results presented in Eqs. (13)-(19) provide important checks of the results presented in the Supplementary Material.  Results for the limit $K'\rightarrow\infty$ holding $K$ constant are equilavent to taking the limit $J'/J\rightarrow\infty$ at finite $T$.  Thus, we conclude that as $J'/J\rightarrow\infty$, the two-layer CCDW behaves  as a one-layer CCDW with a near-neighbor coupling strength $2J$,  exactly twice the coupling  $J$ for the neighboring charges in  a single layer.  In addition, the three-layer  CCDW behaves  as a one-layer CCDW with a near-neighbor coupling strength $3J$,  exactly three times the coupling  $J$ for the neighboring charges in  a single layer. Thus, the CCDW is greatly stabilized in two- and three-layer cuprate superconductors. 

We note that the orbital symmetry of the CCDW in such systems in real space has the $d_{x^2-y^2}$ form. This explains why many workers have been confused in their interpretations of  phase-sensitive experiments on  two-layer cuprates such as Bi2212 and YBCO.  Many of those experiments were measuring the nodal symmetry of the CCDW, not the symmetry of the superconducting order parameter, which is nodeless. In fact, in Bi2212, the superconducting gap function appears to be essentially isotropic,\cite{Zhong} as pictured in Fig. 3, so that  the standard BCS theory applies at least up to 90K. Although while nodeless, the superconducting gap appears to have about a factor of 2 anisotropy in YBCO\cite{Cybart}.  Hence, some forms of the electron-phonon interaction, possibly including optical phonons,  are very likely to provide the primary mechanisms for the superconductivity in the high-temperature cuprates.   However, the nodal CCDW with $d_{x^2-y^2}$-wave symmetry is the likely candidate for enhancing the output power from the coordinated intrinsic Josephson junctions in Bi2212.  Further experimental studies are required to confirm the details of this enhancement.

\section{Supplementary Material}

As  mentioned regarding $Z_{312}$ in the text, 
\begin{eqnarray}
\frac{Z_{312}(K,K')}{Z^3_{112}(K')}&=&\frac{e^{-4K}}{4(1+e^{2K'})^2}\Bigl(1-e^{2K'}+e^{4K'}\nonumber\\
&&+2e^{4K}+4e^{2(K+K')}+2e^{4(K+K')}\nonumber\\
&&+2e^{2(K'+2K)}+e^{4(K'+2K)}\nonumber\\
& &+4e^{2(K'+3K)}-2e^{2(K'+4K)}\Bigr).
\end{eqnarray}
Taking the limit as $K'\rightarrow\infty$ leads to Eq. (13) in the text.

\begin{table}
{\bf Table of points and bonds}\\
\hskip0pt LM1\hskip14pt$n_b$\hskip14pt   LM2\hskip15pt  $n_b$\hskip14pt LM3\hskip15pt $n_b$\\
211\hskip20pt   1\hskip20pt      212\hskip20pt   4\hskip20pt 213\hskip20pt 7\\
\hskip4pt 221\hskip20pt    4\hskip20pt       222\hskip18pt    12\hskip20pt 223\hskip20pt 20\\
 \hskip2pt 311\hskip20pt    2\hskip20pt      312\hskip20pt    7\hskip25pt 313\hskip20pt 12\\
\hskip4pt 321\hskip20pt    7\hskip20pt      322\hskip20pt    20\hskip20pt 323\hskip20pt 33\\
\hskip3pt 331\hskip17pt    12\hskip18pt     332\hskip20pt    33\hskip20pt 333\hskip20pt 54\\
\hskip2pt 411\hskip20pt    3\hskip19pt       412\hskip20pt    10\hskip20pt 413\hskip20pt 17\\
\hskip2pt 421\hskip16pt    10\hskip18pt     422\hskip20pt    28\hskip20pt 423\hskip20pt 46\\
\hskip-60pt 431\hskip16pt    17\hskip18pt     432\hskip20pt    46\hskip50pt\\
\hskip-60pt 441\hskip16pt    24\hskip18pt     442\hskip20pt    64\hskip50pt\\
\hskip4pt 511\hskip18pt    4\hskip20pt       512\hskip20pt    13\hskip20pt 513\hskip20pt 22\\
\hskip2pt 521\hskip16pt    13\hskip18pt     522\hskip20pt    36\hskip20pt 523\hskip20pt 49\\
\hskip-60pt 531\hskip17pt  22\hskip17pt   532\hskip20pt 59\hskip50pt\\
\hskip4pt 611 \hskip15pt    5\hskip21pt       612\hskip20pt   16\hskip20pt 613\hskip20pt 27\\
\hskip-60pt 621\hskip16pt    16\hskip19pt     622\hskip20pt   44\hskip50pt\\
\hskip-60pt 631\hskip16pt    27\hskip19pt 632\hskip20pt   72\hskip50pt\\
\hskip2pt 711\hskip18pt   6\hskip22pt    712\hskip20pt 19\hskip20pt 713\hskip20pt 32\\ 
\hskip2pt 811\hskip18pt   7\hskip22pt    812\hskip20pt 22\hskip20pt 813\hskip20pt 37\\ 
\hskip2pt 911\hskip18pt   8\hskip22pt    912\hskip20pt 25\hskip20pt 913\hskip20pt 42\\ 
\hskip2pt 10\hskip2pt 11\hskip13pt   9\hskip17pt   10\hskip2pt  12\hskip20pt 28\hskip15pt  10\hskip2pt 13\hskip16pt 47\\ 
\end{table}

In the following, we present our results obtained using Mathematica$^{\copyright}$ analytic software.  We have
\begin{eqnarray}
\frac{Z_{712}(K,K')}{[Z_{112}(K')]^{7}}&\begin{array}{c}
\rightarrow\\ K'\rightarrow\infty\end{array}&\cosh^6(2K)
\end{eqnarray}
\begin{eqnarray}
\frac{Z_{711}}{2^7}&=&\cosh^6(K).\label{Z711}
\end{eqnarray}
\begin{eqnarray}
\frac{2^{16}Z_{632}(K,K')}{[Z_{112}(K')]^{12}}&\begin{array}{c}
\rightarrow\\ K'\rightarrow\infty\end{array}&20508\cosh(2K)+17148\cosh(6K)\nonumber\\
&&+12303\cosh(10K)+7732\cosh(14K)\nonumber\\
& &+4248\cosh(18K)+2057\cosh(22K)\nonumber\\
&&+936\cosh(26K)+400\cosh(30K)\nonumber\\
&&+132\cosh(34K)+44\cosh(38K)\nonumber\\
&&+23\cosh(42K)+4\cosh(46K)\nonumber\\
&&+\cosh(54K).\label{Z632}
\end{eqnarray}
\begin{eqnarray}
\frac{Z_{631}(K)}{4}&=&20508\cosh(K)+17148\cosh(3K)\nonumber\\
&&+12303\cosh(5K)+7732\cosh(7K)\nonumber\\
& &+4248\cosh(9K)+2057\cosh(11K)\nonumber\\
&&+936\cosh(13K)+400\cosh(15K)\nonumber\\
&&+132\cosh(17K)+44\cosh(19K)\nonumber\\
&&+23\cosh(21K)+4\cosh(23K)\nonumber\\
&&+\cosh(27K).\label{Z631}
\end{eqnarray}

\begin{eqnarray}
\frac{2^{10}Z_{622}(K,K')}{[Z_{112}(K')]^{12}}&\begin{array}{c}
\rightarrow\\ K'\rightarrow\infty\end{array}&225+360\cosh(4K)\nonumber\\
&&+231\cosh(8K)+132\cosh(12K)\nonumber\\
& &+46\cosh(16K)+20\cosh(20K)\nonumber\\
&&+9\cosh(24K)+\cosh(32K).\label{Z622}
\end{eqnarray}
\begin{eqnarray}
\frac{Z_{621}(K)}{4}&=&225+360\cosh(2K)\nonumber\\
&&+231\cosh(4K)+132\cosh(6K)\nonumber\\
& &+46\cosh(8K)+20\cosh(10K)\nonumber\\
&&+9\cosh(12K)+\cosh(16K).\label{Z621}
\end{eqnarray}
\begin{eqnarray}
\frac{Z_{612}(K,K')}{[Z_{112}(K')]^{5}}&\begin{array}{c}
\rightarrow\\ K'\rightarrow\infty\end{array}&[\cosh(2K)]^5\label{Z612}
\end{eqnarray}
\begin{eqnarray}
\frac{Z_{611}(K)}{2^6}&=&\cosh^5(K)\label{Z611}\\
\end{eqnarray}
\begin{eqnarray}
\frac{2^{13}Z_{532}(K,K')}{[Z_{112}(K')]^{15}}&\begin{array}{c}
\rightarrow\\ K'\rightarrow\infty\end{array}&1456+2619\cosh(4K)\nonumber\\
&&+1906\cosh(8K)+1139\cosh(12K)\nonumber\\
&&+604\cosh(16K)+296\cosh(20K)\nonumber\\
&&+110\cosh(24K)+37\cosh(28K)\nonumber\\
&&+20\cosh(32K)+4\cosh(36K)\nonumber\\
&&+\cosh(44K).\label{Z532}
\end{eqnarray}
\begin{eqnarray}
\frac{Z_{531}(K)}{4}&=&1456+2619\cosh(2K)\nonumber\\
&&+1906\cosh(4K)+1139\cosh(6K)\nonumber\\
&&+604\cosh(8K)+296\cosh(10K)\nonumber\\
&&+110\cosh(12K)+37\cosh(14K)\nonumber\\
&&+20\cosh(16K)+4\cosh(18K)\nonumber\\
&&+\cosh(22K).\label{Z532}
\end{eqnarray}
\begin{eqnarray}
\frac{2^{8}Z_{522}(K,K')}{[Z_{112}(K')]^{10}}&\begin{array}{c}
\rightarrow\\ K'\rightarrow\infty\end{array}&64+100\cosh(4K)+55\cosh(8K)\nonumber\\
&&+26\cosh(12K)+8\cosh(16K)\nonumber\\
& &+2\cosh(20K)+\cosh(24K).\label{Z522}
\end{eqnarray}
\begin{eqnarray}
\frac{Z_{521}(K,K')}{4}&=&64+100\cosh(2K)+55\cosh(4K)\nonumber\\
&&+26\cosh(6K)+8\cosh(8K)\nonumber\\
& &+2\cosh(10K)+\cosh(12K).\label{Z521}
\end{eqnarray}
\begin{eqnarray}
\frac{Z_{512}(K,K')}{[Z_{112}(K')]^{5}}&\begin{array}{c}
\rightarrow\\ K'\rightarrow\infty\end{array}&\cosh^4(2K).
\label{Z512}
\end{eqnarray}
\begin{eqnarray}
\frac{Z_{511}(K)}{2^5}&=&\cosh^4(K).
\label{Z511}
\end{eqnarray}
\begin{eqnarray}
\frac{2^{15}Z_{442}(K,K')}{[Z_{112}(K')]^{16}}&\begin{array}{c}
\rightarrow\\ K'\rightarrow\infty\end{array}&5736+9984\cosh(4K)\nonumber\\
& &+7344\cosh(8K)+4928\cosh(12K)\nonumber\\
& &+2638\cosh(16K)+1216\cosh(20K)\nonumber\\
& &+584\cosh(24K)+224\cosh(28K)\nonumber\\
& &+72\cosh(32K)+32\cosh(36K)\nonumber\\
& &+8\cosh(40K)+2\cosh(48K).\label{Z442}
\end{eqnarray}
\begin{eqnarray}
\frac{Z_{441}(K)}{2}&=&5736+9984\cosh(2K)\nonumber\\
& &+7344\cosh(4K)+4928\cosh(6K)\nonumber\\
& &+2638\cosh(8K)+1216\cosh(10K)\nonumber\\
& &+584\cosh(12K)+224\cosh(14K)\nonumber\\
& &+72\cosh(16K)+32\cosh(18K)\nonumber\\
& &+8\cosh(20K)+2\cosh(24K).\label{Z441}
\end{eqnarray}
\begin{eqnarray}
\frac{2^{11}Z_{432}(K,K')}{[Z_{112}(K')]^{12}}&\begin{array}{c}
\rightarrow\\ K'\rightarrow\infty\end{array}&818\cosh(2K)+588\cosh(6K)\nonumber\\
& &+362\cosh(10K)+172\cosh(14K)\nonumber\\
& &+64\cosh(18K)+34\cosh(22K)\nonumber\\
& &+8\cosh(26K)+2\cosh(34K).\label{Z432}
\end{eqnarray}
\begin{eqnarray}
\frac{Z_{431}(K)}{2}&=&818\cosh(K)+588\cosh(3K)\nonumber\\
& &+362\cosh(5K)+172\cosh(7K)\nonumber\\
& &+64\cosh(9K)+34\cosh(11K)\nonumber\\
& &+8\cosh(13K)+2\cosh(17K).\label{Z431}
\end{eqnarray}
\begin{eqnarray}
\frac{2^{7}Z_{422}(K,K')}{[Z_{112}(K')]^{8}}&\begin{array}{c}
\rightarrow\\ K'\rightarrow\infty\end{array}&40+48\cosh(4K)\nonumber\\
& &+24\cosh(8K)+14\cosh(12K)\nonumber\\
& &+2\cosh(20K).\label{Z422}
\end{eqnarray}
\begin{eqnarray}
\frac{Z_{421}(K)}{2}&=&40+48\cosh(2K)\nonumber\\
& &+24\cosh(4K)+14\cosh(6K)\nonumber\\
&&+2\cosh(10K).\label{Z421}
\end{eqnarray}
\begin{eqnarray}
\frac{2^{3}Z_{412}(K,K')}{[Z_{112}(K')]^{4}}&\begin{array}{c}
\rightarrow\\ K'\rightarrow\infty\end{array}&6\cosh(2K)+2\cosh(6K).\label{Z412}\nonumber\\
\end{eqnarray}
\begin{eqnarray}
\frac{Z_{411}(K)}{2}&=&6\cosh(K)+2\cosh(3K)\nonumber\\
&=&8\cosh^3(K).\label{Z411}
\end{eqnarray}
\begin{eqnarray}
\frac{2^{7}Z_{332}(K,K')}{[Z_{112}(K')]^{9}}&\begin{array}{c}
\rightarrow\\ K'\rightarrow\infty\end{array}&72+96\cosh(4K)\nonumber\\
& &+46\cosh(8K)+32\cosh(12K)\nonumber\\
& &+8\cosh(16K)+2\cosh(24K).\nonumber\\
\label{Z332}
\end{eqnarray}
\begin{eqnarray}
\frac{Z_{331}(K)}{2}&=&72+96\cosh(2K)\nonumber\\
& &+46\cosh(4K)+32\cosh(6K)\nonumber\\
& &+8\cosh(8K)+2\cosh(12K).\label{Z331}
\end{eqnarray}
\begin{eqnarray}
\frac{2^{4}Z_{322}(K,K')}{[Z_{112}(K')]^{6}}&\begin{array}{c}
\rightarrow\\ K'\rightarrow\infty\end{array}&9\cosh(2K)+6\cosh(6K)\nonumber\\
& &+\cosh(14K).\label{Z322}
\end{eqnarray}
\begin{eqnarray}
\frac{Z_{321}(K)}{2}&=&9\cosh(K)+6\cosh(3K)\nonumber\\
& &+\cosh(7K).\label{Z321}
\end{eqnarray}

\begin{eqnarray}
\frac{2^{2}Z_{312}(K,K')}{[Z_{112}(K')]^{3}}&\begin{array}{c}
\rightarrow\\ K'\rightarrow\infty\end{array}&2+2\cosh(4K)\nonumber\\
&=&4\cosh^2(2K).\nonumber\\
\label{Z322}
\end{eqnarray}
\begin{eqnarray}
\frac{Z_{311}(K)}{2}&=&2+2\cosh(2K)=4\cosh^2(K).\label{Z321}
\end{eqnarray}
For the three layer cases, we found
\begin{eqnarray}
\frac{Z_{313}(K,K')}{Z^3_{113}(K')}&=&\frac{16\cosh(6K')\cosh^2(3K)}{[4\cosh(2K')]^3}\nonumber\\
&\begin{array}{c}
\rightarrow\\ K'\rightarrow\pm\infty\end{array}\cosh^2(3K).&
\end{eqnarray}
\begin{eqnarray}
\frac{Z_{413}(K,K')}{Z^4_{113}(K')}&=&\frac{32\cosh(8K')\cosh^3(3K)}{[4\cosh(2K')]^4}\nonumber\\
&\begin{array}{c}
\rightarrow\\ K'\rightarrow\pm\infty\end{array}&\cosh^3(3K).
\end{eqnarray}
\begin{eqnarray}
\frac{Z_{513}(K,K')}{Z^5_{113}(K')}&=&\frac{2^8\cosh(8K')\cosh^2(K')\cosh^4(3K)}{[4\cosh(2K')]^5}\nonumber\\
&\begin{array}{c}
\rightarrow\\ K'\rightarrow\pm\infty\end{array}&\cosh^4(3K).
\end{eqnarray}
\begin{eqnarray}
\frac{Z_{613}(K,K')}{Z^6_{113}(K')}&=&\frac{2^7\cosh(12K')\cosh^5(3K)}{[4\cosh(2K')]^6}\nonumber\\
&\begin{array}{c}
\rightarrow\\ K'\rightarrow\pm\infty\end{array}&\cosh^5(3K).
\end{eqnarray}
\begin{eqnarray}
\frac{Z_{713}(K,K')}{Z^7_{113}(K')}&=&\frac{2^{10}\cosh^2(K')\cosh(12K')\cosh^6(3K)}{[4\cosh(2K')]^7}\nonumber\\
&\begin{array}{c}
\rightarrow\\ K'\rightarrow\pm\infty\end{array}&\cosh^6(3K).
\end{eqnarray}
\begin{eqnarray}
\frac{Z_{813}(K,K')}{Z^8_{113}(K')}&=&\frac{2^{9}\cosh(16K')\cosh^7(3K)}{[4\cosh(2K')]^8}\nonumber\\
&\begin{array}{c}
\rightarrow\\ K'\rightarrow\pm\infty\end{array}&\cosh^7(3K).
\end{eqnarray}
\begin{eqnarray}
\frac{Z_{913}(K,K')}{Z^9_{113}(K')}&=&\frac{2^{12}\cosh^2(K')\cosh(16K')\cosh^8(3K)}{[4\cosh(2K')]^9}\nonumber\\
&\begin{array}{c}
\rightarrow\\ K'\rightarrow\pm\infty\end{array}&\cosh^8(3K).
\end{eqnarray}
\begin{eqnarray}
\frac{Z_{1013}(K,K')}{Z^{10}_{113}(K')}&=&\frac{2^{11}\cosh(20K')\cosh^9(3K)}{[4\cosh(2K')]^{10}}\nonumber\\
&\begin{array}{c}
\rightarrow\\ K'\rightarrow\pm\infty\end{array}&\cosh^9(3K).
\end{eqnarray}